\documentclass[10pt,final,doublecolumn]{IEEEtran}
\usepackage{amsthm}
\usepackage{amsmath}
\usepackage{latexsym}
\usepackage{graphicx}
\usepackage{bbding}
\usepackage{indentfirst}
\usepackage{algorithm,algorithmic}
\usepackage{setspace}
\usepackage{float}
\usepackage{epstopdf}
\usepackage{amssymb}
\usepackage{amsfonts}
\usepackage{enumerate}
\usepackage{multicol}
\usepackage{color}
\usepackage{slashbox}
\usepackage{bm}
\usepackage{amssymb}
\usepackage{stfloats}
\usepackage{epstopdf}
\usepackage{threeparttable}
\usepackage{epstopdf}
\usepackage{threeparttable}
\usepackage{subfigure}
\usepackage{makecell}
\usepackage{cite}

\IEEEoverridecommandlockouts
\allowdisplaybreaks[4]

\begin{document}

\title{Distributed Channel Estimation and Optimization for 6D Movable Antenna: Unveiling Directional Sparsity}

\author{{Xiaodan Shao, \IEEEmembership{Member,~IEEE}, Rui Zhang, \IEEEmembership{Fellow, IEEE}, Qijun Jiang, Jihong Park, \IEEEmembership{Member,~IEEE}, Tony Q. S. Quek, \IEEEmembership{Fellow, IEEE}, and Robert
				Schober, \IEEEmembership{Fellow, IEEE}}
		
		\thanks{This paper was partially submitted to the IEEE International Conference on Communications (ICC), 2025 \cite{icc}.}
		\thanks{X. Shao is with the Institute for Digital Communications, Friedrich-Alexander-University Erlangen-Nurnberg (FAU), 91054
				Erlangen, Germany (e-mail: xiaodan.shao@fau.de).}	
			
	\thanks{R. Zhang is with School of Science and Engineering, Shenzhen Research Institute of Big Data, The Chinese University of Hong Kong, Shenzhen, Guangdong 518172, China. He is also with the Department of Electrical and Computer Engineering, National University of Singapore, Singapore 117583 (e-mail: elezhang@nus.edu.sg). }
	
		 \thanks{Q. Jiang is with the School of Science and Engineering, Chinese University of Hong Kong, Shenzhen, China 518172 (e-mail: qijunjiang@link.cuhk.edu.cn).}
		
		\thanks{J. Park is with the Singapore University of Technology and Design,
				Singapore 487372 (e-mail: jihong\_park@sutd.edu.sg).}
		
		\thanks{T. Q. S. Quek is with the Singapore University of Technology and Design,
				Singapore 487372 (e-mail: tonyquek@sutd.edu.sg).}
		
		\thanks{R. Schober is with the Institute for Digital Communications, Friedrich-Alexander-University Erlangen-Nurnberg (FAU), 91054
				Erlangen, Germany (e-mail: robert.schober@fau.de).}
		}

\maketitle

\begin{abstract}
Six-dimensional movable antenna (6DMA) is an innovative and transformative technology to improve wireless network capacity by adjusting the 3D positions and 3D rotations of antennas/antenna surfaces based on the channel spatial distribution. In order to achieve optimal antenna positions and rotations, the acquisition of statistical channel state information (CSI) is essential for 6DMA systems. However, the existing works on 6DMA have assumed a central processing unit (CPU) to jointly process the signals of all 6DMA surfaces. This inevitably incurs prohibitively high processing cost and latency for channel estimation due to the vast numbers of 6DMA candidate positions/rotations and antenna elements.  
To tackle this issue, we propose a distributed 6DMA processing architecture to reduce the processing complexity of the CPU by equipping each 6DMA surface with a local processing unit (LPU). Furthermore, we unveil for the first time \textbf{\textit{directional sparsity}} property of the 6DMA channels with respect to the distributed users, where each user has significant channel gains only for a (small) subset of 6DMA position-rotation pairs. Based on this property, we propose a practical three-stage protocol for the 6DMA system and corresponding algorithms to conduct statistical CSI acquisition for all 6DMA candidate positions/rotations, 6DMA position/rotation optimization based on statistical CSI, and instantaneous CSI estimation for user data transmission given the optimized 6DMA positions/rotations.
Simulation results show that the proposed channel estimation algorithms achieve higher accuracy than benchmark schemes, while requiring a lower pilot overhead. Furthermore, it is shown that the proposed 6DMA system with statistical CSI-based position and rotation optimization outperforms fixed-position antenna systems and fluid antenna systems with antenna position optimization only in terms of the users' ergodic sum rate, even if the latter systems have perfect instantaneous CSI.
\end{abstract}


\section{Introduction}
As the successor to today's fifth-generation (5G) wireless network, the forthcoming six-generation (6G) wireless network is expected to provide revolutionary advancements, including enormous data speed, improved reliability, minimal latency, extreme connectivity, ubiquitous coverage, and enhanced intelligence and sensing capabilities \cite{wuwen}. To achieve these ambitious goals, the trend is to equip base stations (BSs) and networks with more antennas, evolving from massive multiple-input multiple-output (MIMO) \cite{LA10, ruig} to cell-free massive MIMO \cite{free}, and extremely large-scale MIMO \cite{exl, zeng}. However,
these approaches lead to higher hardware costs, increased energy consumption, and greater complexity, which may not meet the performance needs of future wireless networks. Although low-cost intelligent reflecting surface (IRS) or reconfigurable intelligent surface (RIS)-assisted MIMO systems \cite{hungirs, shaotarget, shaos, proc} have been proposed for performance enhancements, a key drawback of current MIMO systems remains the fixed positioning of their antennas at the BS. With fixed-position antennas (FPAs), the network cannot effectively adapt its resources to varying channel conditions, which limits its ability to adjust to changes in the users' three-dimensional (3D) spatial distribution.

\begin{figure*}[t!]
	\centering
	\setlength{\abovecaptionskip}{0.cm}
	\includegraphics[width=6.9in]{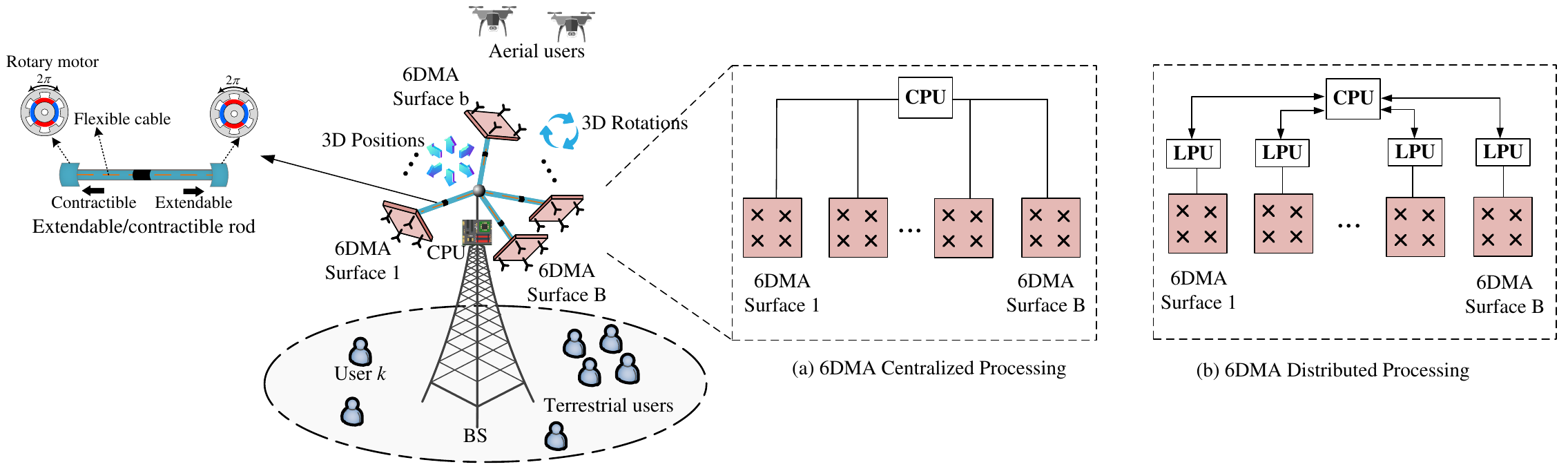}
	\caption{6DMA-equipped BS and different processing architectures.}
	\label{practical_scenario}
\end{figure*}

Recently, to fully exploit the spatial variations of wireless channels
at the transmitter/receiver, six-dimensional movable antenna (6DMA) comprised of multiple rotatable and positionable antennas/antenna surfaces has been proposed as a new technology to improve the performance of 6G wireless networks cost-effectively without the need to deploy additional antennas \cite{shao20246d, 6dma_dis,shaoaga, censing}.
In particular, the integration of 6DMAs into future 6G
wireless networks will fundamentally improve antenna agility and adaptability, and introduce new degrees of freedom (DoFs) for system design.
As shown in Fig. \ref{practical_scenario}, equipped with distributed 6DMA surfaces to match the spatial user distribution, the transmitter/receiver can adaptively allocate antenna resources in space to maximize the array gain and spatial multiplexing gain while also suppressing interference. This thus significantly enhances the capacity of wireless communication networks.
In practice, the positions and rotations of 6DMAs can be adjusted continuously \cite{shao20246d} or discretely \cite{6dma_dis} based on the surface movement mechanism. Moreover, the authors in \cite{censing} proposed a 6DMA-aided wireless sensing system and compared it with FPAs \cite{LA10} and fluid antenna systems, also referred to as two-dimensional (2D) movable antennas \cite{new2023fluid, zhu2}, for both directive and isotropic antenna radiation patterns. Specifically, although 6DMA may incur a moderately higher hardware/energy cost than existing FPAs, it is expected to achieve a much better performance than FPAs by dynamically adapting the antenna positions and rotations to the users' spatial distribution \cite{passive}.
	In addition, compared to fluid antenna systems, which require frequent antenna movement, 6DMA not only significantly improves performance but also requires much less frequent position and/or rotation adjustments. This is because fluid antennas are typically constrained to antenna movement on a 2D surface or along a predefined line with fixed antenna rotations to exploit small-scale channel fading and mitigate deep fades. In contrast, 6DMA achieves its performance gains mainly through the adaptive allocation of antenna resources and spatial DoFs based on the slowly varying spatial distribution of the users, which changes on a minute-by-minute, hourly, or even longer basis in practice. Furthermore, compared with spherical antennas \cite{spha}, where the antenna elements are uniformly distributed over the surface of a sphere, 6DMA offers dynamic adaptability through physical movement, enabling greater spatial flexibility and suppressing interference more effectively with fewer antenna elements.

Although the benefits of 6DMA in wireless communication and sensing
systems have been demonstrated, the channel state information (CSI) acquisition problem for 6DMA still remains largely unexplored in the literature. The performance gains of 6DMA-aided communication heavily rely on the matching of the positions and rotations of the distributed 6DMA surfaces to the users' spatial channel distribution \cite{shao20246d, 6dma_dis}. To obtain optimal 6DMA positions and rotations, knowledge of the CSI of the channels between all 6DMA candidate positions/rotations and all users is essential. However, compared to conventional FPA and fluid antennas, 6DMA channel estimation faces the following new challenges. Firstly, FPA has a small number of channel coefficients to estimate, which is equal to the product of the numbers of transmit and receive antennas.  
In contrast, for 6DMA channels, the CSI in a continuous space with a practically vast number of candidate antenna positions and/or rotations needs to be acquired. Secondly,  for
fluid antennas \cite{new2023fluid,qingmove} in a given 2D space, 
all antenna positions have the same average channel gain for each user, which facilitates CSI acquisition. However, in 6DMA systems, the channels between a user and different candidate antenna positions/rotations in continuous 3D space generally exhibit significantly different distributions (e.g., consider the channels from a user to two 6DMA surfaces whose normal vector points in the 
direction and the opposite direction of the user, respectively). This inevitably makes the CSI estimation problem for 6DMA more intricate and challenging than that for fluid antennas. Due to the above reasons, the methods proposed in the existing literature for channel estimation/acquisition for FPA \cite{ruig,dimen}, and fluid antennas \cite{new2023fluid} are not applicable to 6DMA in general, which motivates this paper.    

On the other hand, the existing works on 6DMA systems \cite{shao20246d, 6dma_dis,censing} have assumed a centralized processing architecture, where a central processing unit (CPU) is connected with all distributed 6DMA surfaces to execute various tasks such as signal precoding/detection, as shown in Fig. \ref{practical_scenario}(a).   
For increasing numbers of antennas and candidate antenna positions/rotations, this centralized processing architecture for 6DMA will have to cope with increasingly higher computational cost and complexity as well as the resulting longer processing delay, which may jeopardize the practical deployment of 6DMA in future wireless networks.  

To reduce the computational burden of the CPU, recent works on FPA-based massive MIMO systems have introduced a decentralized baseband processing architecture \cite{dbp,yanqing,yanqing1}, where the antennas are divided into multiple antenna clusters, and each cluster is equipped with a local baseband processing unit that typically has much lower computational capabilities and thus a lower cost as compared to the traditional CPU.
Inspired by this approach, we propose in this paper a new distributed 6DMA processing architecture, as illustrated in Fig. \ref{practical_scenario}(b), to
alleviate the computational complexity of the CPU. Specifically,  
each 6DMA surface is equipped with a local processing unit (LPU), which carries out various signal processing tasks, such as channel estimation and precoding/combining, independently. All LPUs operate in a decentralized and parallel manner, and they can exchange signals with the CPU for joint signal processing. In this paper, we exploit the proposed 6DMA distributed processing for distributed CSI acquisition at individual 6DMA surfaces, and leverage the CPU for implementing 6DMA position/rotation optimization.
While the distributed 6DMA architecture introduces additional hardware components compared to conventional centralized 6DMA processing with a single CPU, the proposed distributed 6DMA processing utilizing both the CPU and LPUs can not only offload computational tasks from the CPU, but can also reduce the frequency of data exchange between the distributed 6DMA surfaces and the CPU by leveraging the local computation and baseband signal processing capabilities of the LPUs, thus effectively reducing the overall processing cost and latency.         

Different from the prior works on 6DMA systems, which assume known statistical CSI between the BS and the users \cite{shao20246d, 6dma_dis}, this paper addresses the practical CSI acquisition for 6DMA. Particularly, this paper, for the first time, unveils a unique property of 6DMA channels, which we refer to as directional sparsity, i.e., the fact that each user has significant channel gains only with respect to (w.r.t.) a (small) subset of 6DMA position-rotation pairs that can receive dominant direct/reflected signals given the user's location. By exploiting this property, we design new distributed algorithms for estimation of the statistical as well as instantaneous CSI of the users at the 6DMA surfaces. The main contributions of this paper can be summarized as follows. 
\begin{table*}[!t]
	\small
	\caption{Symbol Notations}
	\label{Table1}
	\centering
	\begin{tabular}{|c|c|c|c|c|c|}
		\hline
		Symbol & Description &Symbol & Description \\
		\hline
		$N$&\makecell[c]{Number of antennas of \\ each 6DMA surface}&$M$ &All possible position-rotation pairs \\
		\hline
		$B$&Number of 6DMA surfaces & $\overline{M}$ &Sampling position-rotation pairs \\
		\hline
		$\mathcal{C}$& 6DMA-BS site space & $\mathbf{Z}$& Directional sparsity indicator matrix\\
		\hline
		$\epsilon$&  Directional sparsity detection threshold & $\mathbf{S}$  & Position/rotation selection matrix \\
		\hline
		$\sigma^2$& Average noise power & $\mathbf{s}$ &Position/rotation selection vector\\
		\hline
		$\lambda$ &Carrier wavelength & $L$& Pilot length \\
		\hline
		$p$& Transmit power of user & $K$& Number of users\\
		\hline
		${d}$& \makecell[c]{Minimum antenna spacing\\
			on each 6DMA surface}
		& $\Gamma_{k}$&\makecell[c]{Number of multi-path \\components per user}\\
		\hline
		$\varpi$& \makecell[c]{Regular user ratio}
		& $M_{\mathrm{g}}$& \makecell[c]{Number of position-rotation\\ pairs in each group}\\
		\hline
	\end{tabular}
\end{table*}
\begin{itemize}
\item
First, we present the 6DMA system and channel models as well as the directional sparsity of 6DMA channels, and propose a practical protocol for the operation of a BS equipped with 6DMA surfaces. The proposed protocol consists of three stages. In the first stage, the LPUs independently estimate the statistical CSI of different groups of 6DMA candidate positions and rotations in a parallel manner. In the second stage, based on the estimated statistical CSI, the positions and rotations of all 6DMA surfaces are optimized by the CPU, and then all 6DMA surfaces are moved to their optimized positions and rotations under the joint control of the CPU and their LPUs. In the third stage, with the 6DMA surfaces at their optimized positions and rotations, the instantaneous channels of all the users are estimated at the relevant LPUs in a distributed manner by leveraging the directional sparsity determined in the first stage. It is worth noting that in the above protocol, user data transmission continues in all three stages, where higher rates are achieved in the third stage after the 6DMA surfaces have moved to their optimized positions/rotations \footnote{We note that with this protocol, data transmission by the users takes place continuously with no interruption due to the movement of the 6DMA surfaces. This is because the movement of the surfaces occurs on a much larger time scale compared to the coherence time of the instantaneous user channels. Consequently, the surfaces' positions and rotations can be assumed to be constant within each channel coherence time.}.

\item
Moreover, we propose a joint directional sparsity detection  and channel power estimation algorithm for the statistical CSI acquisition in the first stage, as well as  a
directional sparsity-aided instantaneous channel estimation algorithm in the third stage. Using the estimated average channel power, we further develop a channel power-based optimization algorithm for the second stage to maximize the average sum-rate of all users by jointly optimizing the 3D positions and rotations of all 6DMA surfaces at the BS, subject to practical discrete movement constraints.

\item
Finally, we evaluate the performance of the proposed
statistical CSI and instantaneous CSI estimation algorithms for 6DMA by extensive simulations. Our results demonstrate that the proposed schemes can significantly improve the channel estimation accuracy as compared to several benchmark schemes. Moreover, it is shown that the proposed statistical channel power-based 6DMA position/rotation optimization framework outperforms FPA and fluid antenna systems with 2D antenna position optimization only, even if the latter have perfect instantaneous CSI.
\end{itemize}

The remainder of this paper is organized as follows. Section II presents the 6DMA-BS model with discrete 3D position/rotation adjustment of the 6DMA surfaces and the corresponding channel model, as well as the proposed 6DMA distributed processing architecture.
In Section III, we unveil the directional sparsity of 6DMA channels and propose a practical  protocol for the operation of the 6DMA-BS.
In Sections IV and V, we present the proposed distributed channel estimation algorithms and the proposed position/rotation optimization algorithm, respectively. Section VI provides numerical results and pertinent discussions. Finally, Section VII concludes this paper.

\emph{Notations}: The symbols $(\cdot)^*$, $(\cdot)^H$, and $(\cdot)^T$ represent the conjugate, conjugate transpose, and transpose, respectively. $\mathbb{E}[\cdot]$ is the expectation. $\left \|\cdot\right \|_2$, $\left \|\cdot\right \|_F$, and  $\mathbf{I}_N$ denote the Euclidean norm, the Frobenius norm, and the $N \times N$ identity matrix, respectively. $\|\cdot\|_0$ denotes the $\ell_0$ pseudo-norm, which counts the number of non-zero elements in a vector. \(\mathbf{1}_{N\times 1}\) represents the all-ones vector with dimensions \(N \times 1\). $[\mathbf{a}]_j$ and $[\mathbf{A}]_{i,j}$ denote the $j$-th element of a vector and the $(i,j)$ element of a matrix, respectively. \(\max\{a_1, a_2, \cdots, a_K\}\) represents the largest value in the set. $\text{tr}(\cdot)$ denotes the  trace of a matrix. $\mathcal{O}(\cdot)$ denotes the big-O notation. \(\mathrm{round}(\mathbf{s})\) denotes the operation that rounds each element of  \(\mathbf{s}\) to the nearest integer. $\cup$ denotes the union of two sets. $\boldsymbol{\Gamma}^{\mathrm{c}}$ denotes the complementary set of set $\boldsymbol{\Gamma}$. $\otimes$ denotes the Kronecker product. For ease of reference, the main symbols used in this paper are listed in Table I.

\section{System Model}
\subsection{Distributed 6DMA Architecture}
As illustrated in Fig. \ref{practical_scenario}, the proposed 6DMA-BS comprises $B$ distributed 6DMA surfaces, indexed by set $\mathcal{B} = \{1, 2, \ldots, B\}$. Each surface is modeled as a uniform planar array (UPA) with $N \geq 1$ antennas, indexed by set $\mathcal{N} = \{1, 2, \ldots, N\}$. Each 6DMA surface is equipped with a local computing hardware, i.e., LPU, which carries out the necessary baseband processing tasks, such as channel estimation and signal precoding/combining, in a decentralized and parallel fashion (see Fig. \ref{practical_scenario}(b)). 
Each LPU is connected to the common CPU at the
BS via a separate rod, which houses flexible cables (e.g., coaxial cables). These cables supply power to the 6DMA surface and LPU, and also facilitate control and signal exchange between the LPU and CPU. Moreover, the CPU controls two rotary motors that are mounted at both ends of the rod to adjust the position and rotation of each 6DMA surface. In addition, each rod can flexibly contract and
extend to control the distance between each 6DMA surface and
the CPU \footnote{The contraction and extension of the rods are designed to adjust the spatial configuration of the 6DMA surfaces without altering their electromagnetic properties. Proper shielding and cable design ensure stable signal transmission  regardless of the rod length adjustments.}. For this setup, the BS can function akin to
a “transformer”, possessing the capability to instantly reconfigure its antenna array into virtually any conceivable shape for optimizing wireless network performance. 

We note that while continuously tuning the position and/or rotation of each 6DMA
surface offers the greatest flexibility and thus leads to the highest performance gains over FPAs, it is difficult
to realize in practice since 6DMA surfaces need to be mechanically moved by devices such as rotary motors, which can only adjust the position/rotation in discrete steps. Moreover, continuous adjustments entail high hardware costs, increased power consumption, and significant movement time overhead. Therefore, in this paper, we adopt a discrete 6DMA movement model, which is more cost-effective and practically viable to implement, as the number of discrete positions and rotations can be used to balance
the achievable performance and the implementation cost. Furthermore, for simplicity of implementation, we assume that for each candidate position, there is only one possible rotation\footnote{The proposed framework can be readily extended to the case where for each position, there are multiple rotations \cite{6dma_dis}.}. Specifically, we assume that there are in total $M\geq B$ discrete position-rotation pairs $(\mathbf{q}_m, \mathbf{u}_m)$, denoted by set $\mathcal{M}=\{1,2,\cdots,M\}$, where $\mathbf{q}_m\in \mathbb{R}^3, m\in \mathcal{M}$, and $\mathbf{u}_m\in \mathbb{R}^3, m\in \mathcal{M}$, denote the $m$-th possible discrete position and rotation of the 6DMA surfaces, respectively. They are specified as follows:
\begin{align}\label{bb}
\mathbf{q}_{m}&=[x_{m},y
_{m},z_{m}]^T\in\mathcal{C},\\
\mathbf{u}_{m}&=[\alpha_{m},\beta_{m},\gamma_{m}]^T.
\end{align}
Here \(x_{m}\), \(y_{m}\), and \(z_{m}\) are the coordinates of the 6DMA surface's center at the $m$-th discrete position in a global Cartesian coordinate system (CCS) with origin \(o\), where the CPU is located. The rotation angles \(\alpha_{m}\), \(\beta_{m}\), and \(\gamma_{m}\) correspond to rotations around the $x$-axis, $y$-axis, and $z$-axis, respectively. The space \(\mathcal{C}\) defines the 3D area at the BS site where the 6DMA surfaces can be dynamically positioned and rotated.

\begin{figure}[t]
	\centering
	\setlength{\abovecaptionskip}{0.cm}
	\includegraphics[width=3.3in]{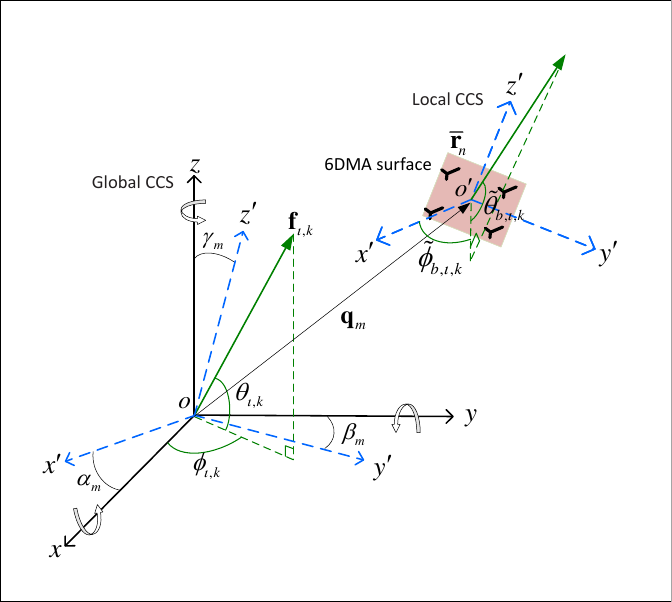}
	\caption{Illustration of the geometry of the 6DMA surface at the $m$-th candidate position-rotation pair.}
	\label{system}
\end{figure}

The rotation matrix $\mathbf{R}(\mathbf{u}_{m})$ for the angles in $\mathbf{u}_{m}$ is given by
\begin{align}\label{R}
	\scalebox{0.8}{$ %
		\begin{aligned}
			\mathbf{R}(\mathbf{u}_{m}) =
			&\begin{bmatrix}
				c_{\beta_{m}}c_{\gamma_{m}} & c_{\beta_{m}}\omega_{\gamma_{m}} & -\omega_{\beta_{m}} \\
				\omega_{\beta_{m}}\omega_{\alpha_{m}}c_{\gamma_{m}}-c_{\alpha_{m}}
				\omega_{\gamma_{m}} & \omega_{\beta_{m}}\omega_{\alpha_{m}}\omega_{\gamma_{m}}+c_{\alpha_{m}}c_{\gamma_{m}} & c_{\beta_{m}}\omega_{\alpha_{m}} \\
				c_{\alpha_{m}}\omega_{\beta_{m}}c_{\gamma_{m}}+\omega_{\alpha_{m}}\omega_{\gamma_{m}} & c_{\alpha_{m}}\omega_{\beta_{m}}\omega_{\gamma_{m}}-\omega_{\alpha_{m}}c_{\gamma_{m}} &c_{\alpha_{m}}c_{\beta_{m}} \\
			\end{bmatrix},
		\end{aligned}$}
\end{align}
where $c_{x}=\cos(x)$ and $\omega_{x}=\sin(x)$ \cite{rot3}. We define a local CCS for the 6DMA surface. As shown in Fig. \ref{system}, each 6DMA surface's local CCS is denoted by $o'\text{-}x'y'z'$, with the surface's center serving as the origin $o'$. The $x'$-axis is oriented along the direction of the normal vector of the 6DMA surface.
The position of the $n$-th antenna on the 6DMA surface in its local CCS is given by $\bar{\mathbf{r}}_{n}\in \mathbb{R}^3$. The rotation $\mathbf{u}_{m}$ and position $\mathbf{q}_{m}$ are then used to find the position of the $n$-th antenna on the 6DMA surface at the $m$-th discrete position in the global CCS. This global position is given by $\mathbf{r}_{m,n}\in \mathbb{R}^3$ as follows: 
\begin{align}\label{nwq}
\!\!\!\!\mathbf{r}_{m,n}(\mathbf{q}_{m},\mathbf{u}_{m})=\mathbf{q}_{m}+\mathbf{R}
(\mathbf{u}_{m})\bar{\mathbf{r}}_{n},~n\in\mathcal{N},~m \in\mathcal{M}.
\end{align}

We define \(\mathcal{Q} = \{(\mathbf{q}_1, \mathbf{u}_1), (\mathbf{q}_2, \mathbf{u}_2), \ldots, (\mathbf{q}_M, \mathbf{u}_M)\}\) as the set of $M$ possible discrete position-rotation pairs for the 6DMA surfaces \footnote{In a practical system, choosing \( M \) requires a comprehensive evaluation of the trade-offs between resource cost and system performance.}. Let \(i_b \in \mathcal{M}\) be the index of the chosen position-rotation pair for the $b$-th 6DMA surface, \(b \in \mathcal{B}\). Thus, the position-rotation pair for the $b$-th surface is \((\mathbf{q}_{i_b}, \mathbf{u}_{i_b}) \in \mathcal{Q}\). As explained in \cite{shao20246d,6dma_dis}, the following three practical constraints for rotating/positioning 6DMA surfaces need to be considered:
\begin{align}
&\mathbf{n}(\mathbf{u}_{i_b})^T(\mathbf{q}_{i_j}-\mathbf{q}_{i_b})\leq  0,~\forall b ,j \in \mathcal{B}, j\neq b, \label{rcc}\\
&\mathbf{n}(\mathbf{u}_{i_b})^T\mathbf{q}_{i_b}\geq 0,~\forall b\in \mathcal{B}, \label{dd}\\
&\|\mathbf{q}_{i_b}-
\mathbf{q}_{i_j}\|_2\geq d_{\min},~\forall b ,j \in \mathcal{B}, j\neq b, \label{dss}
\end{align}
where vector \(\mathbf{n}(\mathbf{u}_{i_b}) = \mathbf{R}(\mathbf{u}_{i_b})\bar{\mathbf{n}}\) represents the normal vector of the $b$-th 6DMA surface in the global CCS, with \(\bar{\mathbf{n}}\) being the normal vector in the local CCS. Constraint \eqref{rcc} prevents mutual signal reflections between surfaces, as it ensures that a 6DMA surface does not form an acute angle with any of the other 6DMA surfaces. Constraint \eqref{dd} ensures that the CPU does not block the signal of
the 6DMA surfaces, which is achieved by tuning the normal vectors of the 6DMA surfaces such that they do not point towards the CPU. Finally, constraint \eqref{dss} guarantees
a minimum distance, $d_{\min}$, between any two 6DMA surfaces
to prevent overlapping and coupling.

\subsection{Channel Model}
We focus on uplink multiuser transmission, where \( K \) users, each equipped with a single FPA, are spatially distributed throughout the cell. Assuming the presence of a multipath channel between each user and the BS, the far-field channel from user \( k \) to the \( B \) 6DMA surfaces, denoted by \(\mathbf{h}_k(\mathbf{q},\mathbf{u})\in \mathbb{C}^{NB\times 1}\), is defined as follows:

\begin{align}\label{vg}
	\mathbf{h}_{k}(\mathbf{q},\mathbf{u})&=
	[\mathbf{h}_{1,k}^T(\mathbf{q}_{i_{1}},\mathbf{u}_{i_{1}}),\cdots, \mathbf{h}_{B,k}^T(\mathbf{q}_{i_{B}},\mathbf{u}_{i_{B}})]^T,
\end{align}
with
\begin{align}
	{\mathbf{q}}&=[\mathbf{q}_{i_1}^T,\mathbf{q}_{i_2}^T,\cdots,\mathbf{q}_{i_B}
	^T]^T\in \mathbb{R}^{3B\times 1},\label{lk}\\
	\mathbf{u}&=[\mathbf{u}_{i_1}^T,\mathbf{u}_{i_2}^T,\cdots,\mathbf{u}_{i_B}^T]^T\in \mathbb{R}^{3B\times 1} \label{lk1}.
\end{align}
In \eqref{vg}, $\mathbf{h}_{b,k}(\mathbf{q}_{i_{b}},\mathbf{u}_{i_{b}})\in \mathbb{C}^{N\times 1}$ represents the channel from the $k$-th user to all the antennas of the $b$-th 6DMA surface at the 6DMA-BS, and can be expressed as
\begin{align}\label{uk}
\mathbf{h}_{b,k}(\mathbf{q}_{i_{b}},\mathbf{u}_{i_{b}})&=
\sum_{\iota=1}^{\Gamma_{k}}e^{-j\varphi_{\iota, k}}\sqrt{\mu_{\iota,k}}
\sqrt{g_{\iota,k}(\mathbf{u}_{i_{b}})}\mathbf{a}_{b,\iota,k}
(\mathbf{q}_{i_{b}},\mathbf{u}_{i_{b}}),
\end{align}
where $\Gamma_{k}$ represents the total number of channel paths from user \( k \) to the BS, and ${\mu}_{\iota, k}$ and $\varphi_{\iota, k}$ denote the path gain and
phase shift from user $k$ to the BS along path \( \iota \), respectively. $\mathbf{a}_{b,\iota,k}
(\mathbf{q}_{i_{b}},\mathbf{u}_{i_{b}})$ and $g_{\iota,k}(\mathbf{u}_{i_{b}})$ denote the 6D steering vector and the effective antenna gain of the $b$-th 6DMA surface, respectively, as defined below. 

First, the 6D steering vector of the \( b \)-th 6DMA surface at the BS for receiving a signal from user \( k \) over path \( \iota \) is expressed as 
\begin{align}\label{gen}
&\mathbf{a}_{b,\iota,k}(\mathbf{q}_{i_{b}},\mathbf{u}_{i_{b}})\nonumber\\
&= \!\left[\!e^{-j\frac{2\pi}{\lambda}
\mathbf{f}_{\iota,k}^T\mathbf{r}_{i_b,1}(\!\mathbf{q}_{i_{b}},
\mathbf{u}_{i_{b}}\!)},
\!\cdots,\! e^{-j\frac{2\pi}{\lambda}\mathbf{f}_{\iota,k}^T
\mathbf{r}_{i_b,N}(\!\mathbf{q}_{i_{b}},\mathbf{u}_{i_{b}}\!)}\!\right]^T,
\end{align}
where $\lambda$ denotes the carrier wavelength, and $\mathbf{f}_{\iota,k}$ represents the direction-of-arrival (DOA) vector corresponding to direction $(\theta_{\iota,k}, \phi_{\iota,k})$ in the global CCS, which is defined as
\begin{align}\label{KM}
\!\!\!\mathbf{f}_{\iota,k}\!=\![\cos(\theta_{\iota,k})\cos(\phi_{\iota,k}), \cos(\theta_{\iota,k})\sin(\phi_{\iota,k}), \sin(\theta_{\iota,k})]^T,
\end{align}
where the azimuth and elevation angles for the \(\iota\)-th channel path between user \(k\) and the BS are given by \(\phi_{\iota,k}\in[-\pi,\pi]\) and \(\theta_{\iota,k}\in[-\pi/2,\pi/2]\), respectively. 

Next, to conveniently define $g_{\iota,k}(\mathbf{u}_{i_{b}})$, we project the DOA vector \( \mathbf{f}_{\iota,k} \) onto the local CCS of the \( b \)-th 6DMA surface, denoted by \( \tilde{\mathbf{f}}_{b,\iota,k} \), as
\begin{align}
	\tilde{\mathbf{f}}_{b,\iota,k}=-\mathbf{R}(\mathbf{u}_{i_{b}})^{-1}\mathbf{f}_{\iota,k}.
\end{align}
Then, we represent $\tilde{\mathbf{f}}_{b,\iota,k}$ in the spherical coordinate system as
\begin{align}\label{KM3}
&\tilde{\mathbf{f}}_{b,\iota,k}\nonumber\\
&=[\cos(\tilde{\theta}_{b,\iota,k})\cos(\tilde{\phi}_{b,\iota,k}), \cos(\tilde{\theta}_{b,\iota,k})\sin(\tilde{\phi}_{b,\iota,k}), \sin(\tilde{\theta}_{b,\iota,k})]^T,
\end{align}
where
$\tilde{\theta}_{b,\iota,k}$ and $\tilde{\phi}_{b,\iota,k}$ represent the corresponding  DOAs in the local CCS (see Fig. \ref{system}).
Finally, the effective antenna gain $g_{\iota,k}(\mathbf{u}_{i_{b}})$ of the \(b\)-th 6DMA surface along direction \((\tilde{\theta}_{b,\iota,k}, \tilde{\phi}_{b,\iota,k})\) in the linear scale can be defined as 
\begin{align}\label{gm}
g_{\iota,k}(\mathbf{u}_{i_{b}})=10^{\frac{A(\tilde{\theta}_{b,\iota,k}, \tilde{\phi}_{b,\iota,k})}{10}},~b\in\mathcal{B}, \iota\in\Gamma_{k}, k\in\mathcal{K},
\end{align}
where \(A(\tilde{\theta}_{b,\iota,k}, \tilde{\phi}_{b,\iota,k})\) denotes the effective antenna gain in dBi, which is determined by the radiation pattern of the selected antenna \cite{3gpp, yumeng,shao20246d}.

It is worth noting that the 6DMA channel in \eqref{uk} is a function
of the positions and rotations of all 6DMA surfaces, indicating that the channel quality can be adjusted by positioning and rotating the 6DMA surfaces. To reconfigure the positions and rotations of the 6DMA surfaces and obtain a desired channel quality, accurate statistical CSI between
all the users and the $M$ 6DMA candidate position-rotation pairs is required.
\begin{figure*}[t!]
	\centering
	\setlength{\abovecaptionskip}{0.cm}
	\includegraphics[width=6.3in]{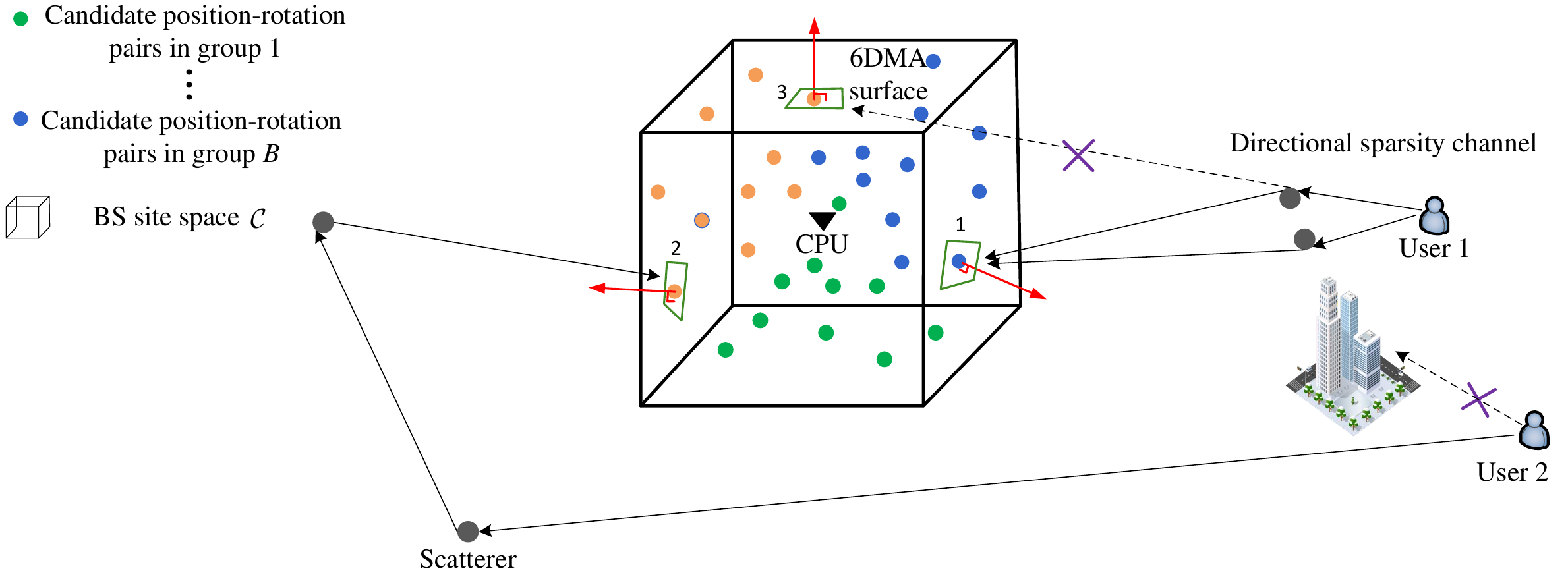}
	\caption{Illustration of the directional sparsity of 6DMA channels and the 6DMA candidate position-rotation pair grouping.}
	\label{candidate}
\end{figure*} 
\subsection{Achievable Sum Rate Analysis}
The multiple-access channel from the $K$ users to a 6DMA surface at all $M$ possible discrete positions and rotations is characterized by $\mathbf{H}=[\mathbf{H}_{1}^T,\cdots,\mathbf{H}_{M}^T]^T
\in\mathbb{C}^{MN\times K}$ with 
\begin{align}
	\mathbf{H}_m=[\mathbf{h}_{m,1}(\mathbf{q}_{m},\mathbf{u}_{m}),\cdots,\mathbf{h}_{m,K}(\mathbf{q}_{m},\mathbf{u}_{m})]&\in \mathbb{C}^{N\times K}, 
\end{align}
denoting the channel from all $K$ users to all the antennas of the $m$-th 6DMA candidate position-rotation pair. In addition, we denote 
\begin{align}
	\bar{\mathbf{H}}(\mathbf{q},\mathbf{u})=[\mathbf{h}_1(\mathbf{q},\mathbf{u}),\mathbf{h}_2(\mathbf{q},\mathbf{u}),\cdots,
	\mathbf{h}_{K}(\mathbf{q},\mathbf{u})]\in \mathbb{C}^{BN\times K},
\end{align}
as the multiple-access channel from all $K$ users to all $B$ 6DMA surfaces at the BS assuming each surface has selected one fixed position-rotation pair. It can be shown that $\bar{\mathbf{H}}(\mathbf{q},\mathbf{u})$ can be expressed in terms of $\mathbf{H}$ as
\begin{align}
	\bar{\mathbf{H}}(\mathbf{q},\mathbf{u})=(\mathbf{S} \otimes \mathbf{I}_N) \mathbf{H},
\end{align}
where $\mathbf{S}\in\mathbb{R}^{B\times M}$ is the position/rotation selection matrix, which is defined by
\begin{align}\label{3b}
	[\mathbf{S}]_{b,m}= \left\{\begin{matrix}
		1, & \text{if the $b$-th 6DMA surface selects candidate} \\
			&\text{position-rotation pair $m$},\\
		0, & \text{otherwise},
	\end{matrix}\right.
\end{align}
for $b\in\mathcal{B}$ and $m\in\mathcal{M}$.

Then, the received signals at the BS are given by
\begin{align}\label{ly}
	\mathbf{y}&=(\mathbf{S} \otimes \mathbf{I}_N)\mathbf{H}
	\bar{\mathbf{x}}+\mathbf{w},
\end{align}
where $\mathbf{w}\sim\mathcal{CN}(\mathbf{0}_{NB},\sigma^2\mathbf{I}_{NB})$ denotes the complex additive white Gaussian noise (AWGN) vector at the BS with zero mean and average power $\sigma^2$.
In \eqref{ly}, $\bar{\mathbf{x}}=\sqrt{p }[\bar{x}_1, \bar{x}_2,\cdots, \bar{x}_{K}]^T\in \mathbb{C}^{K\times 1}$, where $\bar{x}_k$ is the normalized transmit signal of user $k$ with unit average power, and $p$ denotes the transmit power of each user.

Based on \eqref{ly} and by applying the expectation w.r.t. the random channel $\mathbf{H}$, the achievable ergodic sum rate of the users 
is given by \cite{david}
\begin{subequations}
\begin{align}
	&C(\mathbf{s})=\mathbb{E}\left[\log_2 \det \left(\mathbf{I}_{K}+\frac{p}{\sigma^2}\mathbf{H}^H
	(\mathbf{S} \otimes \mathbf{I}_N)^T(\mathbf{S} \otimes \mathbf{I}_N)
	\mathbf{H}\right)\right]\label{qaaa1}\\
	&=\mathbb{E}\left[\log_2 \det \left(\mathbf{I}_{K}+\frac{p}{\sigma^2}\mathbf{H}^H
	(\text{diag}(\mathbf{s})\otimes \mathbf{I}_N)
	\mathbf{H}\right)\right],\label{qaaa}
\end{align}
\end{subequations}
where \eqref{qaaa} holds as \( \mathbf{S}^T\mathbf{S} = \text{diag}(\mathbf{s}) \) and $(\mathbf{S} \otimes \mathbf{I}_N)^T(\mathbf{S} \otimes \mathbf{I}_N)=\text{diag}(\mathbf{s})\otimes \mathbf{I}_N$, which simplifies the representation by reducing the high-dimensional position/rotation selection matrix \( \mathbf{S} \) to the lower-dimensional position/rotation selection vector
\(\mathbf{s} \in \mathbb{R}^{M \times 1}\), defined as 
\begin{align}\label{3b}
	\!\![\mathbf{s}]_m= \left\{\begin{matrix}
		1, & \text{if position-rotation pair}~m~\text{is selected}\\
		& \text{for a 6DMA surface},\\
		0, &\mathrm{otherwise},
	\end{matrix}\right.
\end{align}
for $m\in\mathcal{M}$. Note that the ergodic sum rate in \eqref{qaaa} depends only on which position-rotation pairs are selected, but not on their assignment to specific individual surfaces. This is possible since all surfaces have identical properties and simplifies the design process by focusing on the selection of optimal pairs.

The exact ergodic sum rate in \eqref{qaaa} is hard to obtain,
hence we resort to deriving an upper bound for it by exploiting Jensen's inequality, which will be used in the subsequent optimization. Assuming that the channels of all users are statistically independent, an upper bound for $C(\mathbf{s})$ in \eqref{qaaa} can be obtained as follows
\begin{subequations}
\begin{align} 
	&C(\mathbf{s})\leq \bar{C}(\mathbf{s})\nonumber\\
	&=\log_2 \det \left(\mathbf{I}_{K}+\frac{p}{\sigma^2}\mathbb{E}\left[\mathbf{H}^H
	(\text{diag}(\mathbf{s})\otimes \mathbf{I}_N)
	\mathbf{H}\right]\right)
\label{o4}\\
	&=\log_2 \det \left(\mathbf{I}_{K}+\frac{p}{\sigma^2}\mathbb{E}
	\left[\sum_{m=1}^{M}[\mathbf{s}]_m\sum_{j=N(m-1)+1}^{Nm}[\mathbf{H}]_{j,:}^H[\mathbf{H}]_{j,:}\right]\right)	\label{o3}\\
	&=\sum_{k=1}^K\log_2 \left( 1+\frac{p}{\sigma^2}\sum_{m=1}^{M}[\mathbf{s}]_m\sum_{j=N(m-1)+1}^{Nm} \mathbb{E}\left[|[\mathbf{H}]_{j,k}|^2\right]\right)	\label{o2}\\
	&=\sum_{k=1}^K\log_2 \left( 1+\frac{p}{\sigma^2}\sum_{m=1}^{M}[\mathbf{s}]_m[\mathbf{P}]_{m,k}\right),\label{o1}
\end{align}
\end{subequations}
where \eqref{o1} holds due to the assumed independence of the channels of different users, and
\(\mathbf{P} \in \mathbb{R}^{M \times K}\) represents the average channel power matrix, with its \((m,k)\)-th element, \(\sum_{j=N(m-1)+1}^{Nm} \mathbb{E}\left[|[\mathbf{H}]_{j,k}|^2\right]\), denoting the average power of the channels between user \(k\) and all antennas of a 6DMA surface located at candidate position-rotation pair \(m\). 

It is worth noting that different from the conventional multiuser channel obtained for FPAs, the ergodic sum rate for 6DMAs depends on the candidate position/rotation statistical
CSI (i.e., the average channel power), $\mathbf{P}$, and position/rotation selection vector $\mathbf{s}$,
which underlines the importance of estimating the candidate position/rotations' statistical CSI $\mathbf{P}$, as only when $\mathbf{P}$ is known, the position/rotation selection $\mathbf{s}$ can be optimized to maximize the ergodic sum rate. 

\section{Directional Sparsity and Protocol Design}
In this section, we introduce the peculiar directional sparsity property of 6DMA channels and propose a practical protocol for the operation of the 6DMA-BS based on distributed signal processing. 

\subsection{Directional Sparsity of 6DMA Channels}
\begin{figure}[t!]
	\centering
	\subfigbottomskip=2pt
	\subfigcapskip=-5pt
	\subfigure[Uniformly distributed users (only regular users
	are present in the network, see Section VI for details).]{
		\includegraphics[width=0.80\linewidth]{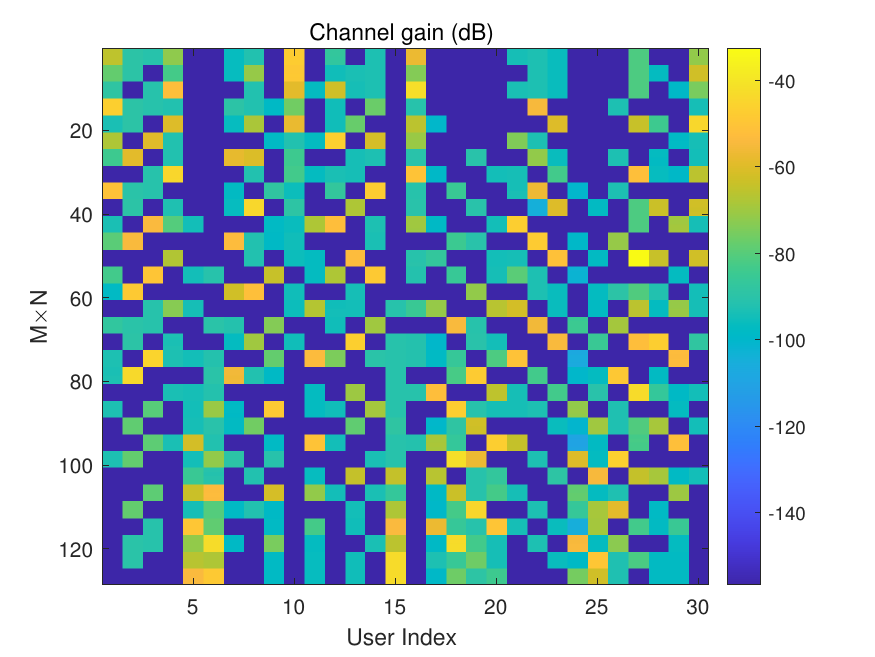}}
	\subfigure[Non-uniformly distributed users (two types of users are present in the network, namely regular users
	and hotspot users, with the proportion of regular users given by $ \varpi = \frac{\text{number of regular users}}{\text{total number of users}} = 0.3$, see Section VI for details).]{
		\includegraphics[width=0.80\linewidth]{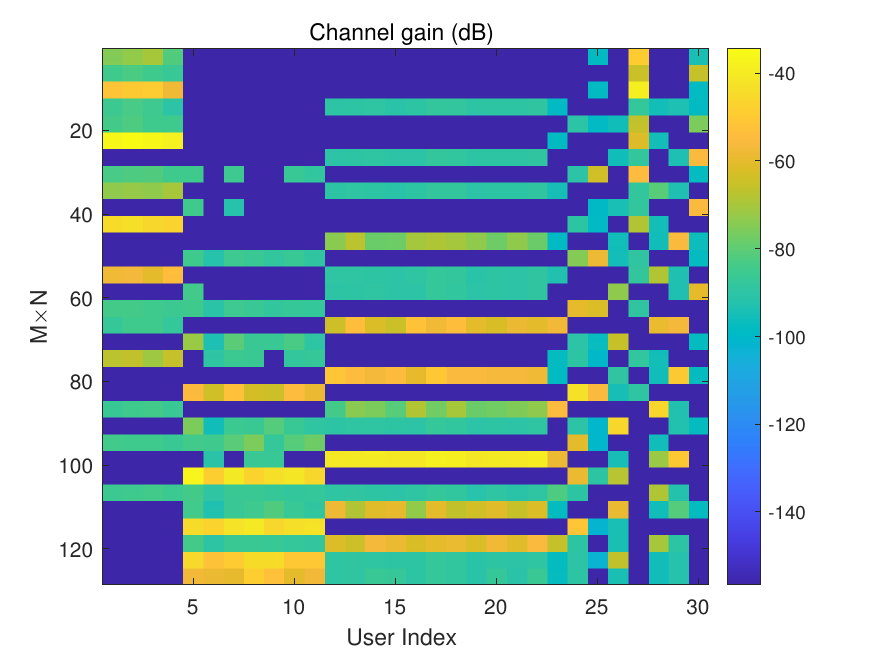}}
	\caption{An illustration of the sparse pattern in $\mathbf{H}$.}
	\label{gain_N}
\end{figure}

\begin{figure*}[t!]
	\centering
	\setlength{\abovecaptionskip}{0.cm}
	\includegraphics[width=6.9in]{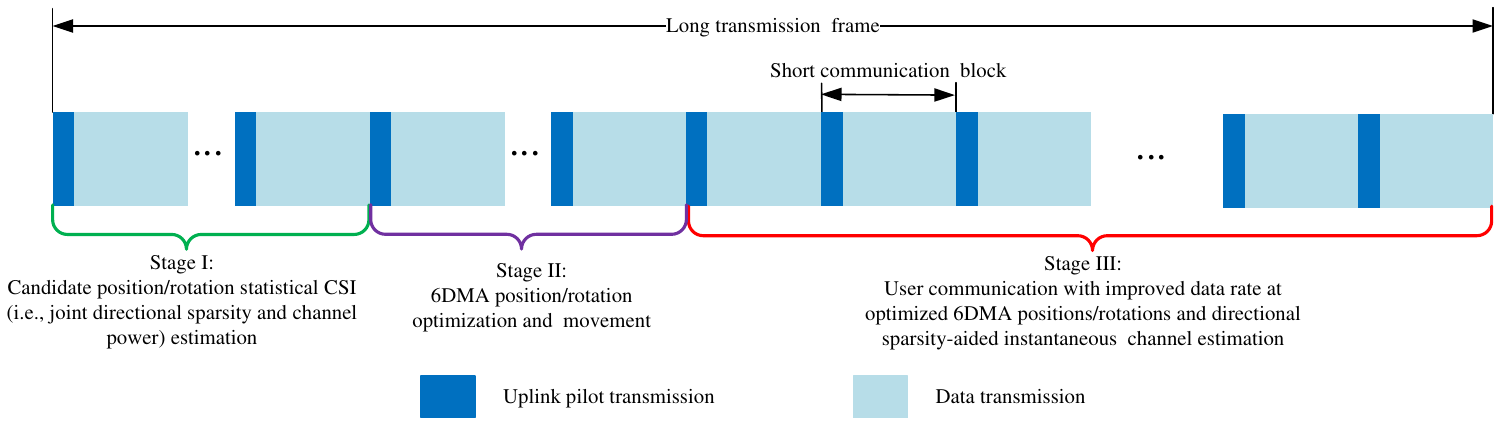}
	\caption{The proposed three-stage protocol for 6DMA system.}
	\label{timescale}
\end{figure*}

In existing wireless networks, the BS is usually installed at high altitude. Macro BSs, for instance, are generally installed at heights ranging from 30 to 70 meters \cite{bsh}. Thus, even in typical urban  environments with multipath scattering, most scatterers will be located near the ground-level users, while there will be only a limited number of scatterers around the BS \cite{bsh}. Due to the rotatability, positionability, and antenna directivity of 6DMA, the channels between a given user and different candidate 6DMA positions and rotations in a continuous 3D space generally exhibit drastically different distributions.
In particular, each user $k$ may have channels with significant gains only w.r.t. a small subset of 6DMA position-rotation pairs indexed by the set $\mathcal{W}_k \subseteq \mathcal{M}$. While for the remaining candidate position-rotation pairs $m \in \mathcal{W}_k^{\mathrm{c}}$, which may either face in the opposite direction of the user or be blocked by obstacles, the corresponding channels of the user may be much weaker and thus can be ignored. In the above, we have $
\mathcal{W}_k \cup \mathcal{W}_k^{\mathrm{c}} = \mathcal{M}$ and $ \mathcal{W}_k \cap \mathcal{W}_k^{\mathrm{c}} = \emptyset
$. For example, for the case considered in Fig. \ref{candidate}, user 1 can establish a significant channel with candidate position-rotation pair 1, but its channels with candidate position-rotation pairs 2 and 3 are much weaker and thus can be assumed to be approximately zero. Similarly, in Fig. \ref{practical_scenario}, user $k$ on the ground 
can establish a significant channel with 6DMA surface 1, but its
channel with 6DMA surface $b$, which is directed towards the sky, is much weaker and thus can  be
assumed to be approximately zero.
These examples illustrate that sparsity exists in the channels between the users and different 6DMA position-rotation pairs, which we define as \textbf{\textit{directional sparsity}} as follows.

{\textbf{Definition 1 (Directional Sparsity)}}:
In the considered 6DMA channel model in \eqref{uk}, user $k$ is assumed to have positive channel gains to only a subset of 6DMA position-rotation pairs $m\in \mathcal{W}_k$  
(i.e., the corresponding antenna gain $g_{\iota,k}(\mathbf{u}_{m})\neq 0, m\in \mathcal{W}_k$, in \eqref{uk}); while for the remaining 6DMA position-rotation pairs $m\in \mathcal{W}_k^{\mathrm{c}}$, the channels to user $k$ are assumed to be zero (i.e.,  $g_{\iota,k}(\mathbf{u}_{m})=0, m\in \mathcal{W}_k^{\mathrm{c}}$, in \eqref{uk}). 

An illustration of the directional sparsity of the 6DMA channels from the \(K\) users to all 6DMA discrete position-rotation pairs, i.e., \(\mathbf{H} \in \mathbb{C}^{MN \times K}\), is provided in Fig. \ref{gain_N} \footnote{Note that the directional sparsity applies not only to \(\mathbf{H} \in \mathbb{C}^{MN \times K}\), but also to the 6DMA channel \(\bar{\mathbf{H}}(\mathbf{q}, \mathbf{u}) \in \mathbb{C}^{BN \times K}\), where the $B$ 6DMA surfaces select $B$ out of the $M$ candidate position-rotation pairs.}. It is interesting to observe that the channel gains of $\mathbf{H}$ exhibit a `{\it block sparsity}' pattern, for both the cases of uniform and non-uniform user distributions. This is due to the directional sparsity for each user's channels to all candidate position-rotation pairs as well as the fact that all $N$ antennas of a 6DMA surface for a given position-rotation pair have the same channel power distribution. 

To characterize the directional sparsity of 6DMA channels, we define $\mathbf{Z}\in \mathbb{R}^{M \times K}$ as the directional sparsity indicator matrix with $[\mathbf{Z}]_{m,k} = 1$ if the $k$-th user w.r.t. a 6DMA surface located at the $m$-th candidate position-rotation pair has a positive channel power, and $[\mathbf{Z}]_{m,k} = 0$ otherwise. Thus, we have
\begin{align}
	[\mathbf{Z}]_{m,k} = \begin{cases} 
		1, & \text{if the channel between the}~ k \text{-th user} \\
		&\text{and the}~ m \text{-th candidate 6DMA}\\ &  \text{ position-rotation is non-zero}, \\
		0, & \text{otherwise}.
	\end{cases}
\end{align}

It is worth noting that the directional sparsity in the context of 6DMA is different from the traditional concept of angular channel sparsity in MIMO systems with FPAs \cite{ruig}, which generally refers to the scenario where the number of dominant channel paths from each user is significantly smaller than the number of FPAs at the transmitter/receiver, regardless of their positions/rotations.      

\subsection{Protocol Design}
Next, we propose a practical three-stage protocol for the operation
of the 6DMA-BS, as
illustrated in Fig. \ref{timescale}. In Stage I, the 6DMA surfaces are moved to some sample position/rotation pairs, aiming to estimate the statistical CSI of all candidate position/rotation within a long transmission frame, i.e., over a long period of time during which the statistical CSI is assumed to be approximately constant. In Stage II, the positions and rotations of the 6DMA surfaces are optimized based on the estimated statistical CSI and then the surfaces are moved to the optimized locations. In Stage III, since the instantaneous channel for the optimized positions/rotations is time-varying due to small-scale channel fading, we need to estimate it for all users in each  short communication block \footnote{Note that the instantaneous channels between the users and the
		$B$ 6DMA surfaces in Stage I and Stage II are also estimated for data communication by using a traditional unlink channel estimation method \cite{LA10}.} (see Fig. \ref{timescale}). Notice that the movement of the 6DMA surfaces in Stages I and II occurs over a much larger time scale compared to the channel coherence time, which ensures that this movement neither affects the accuracy of channel estimation nor interrupts the continuous user data transmission. Specifically, the three stages of the proposed protocol are explained in detail. 
\begin{itemize}
	\item Stage I (Candidate position/rotation statistical CSI
	(i.e., joint directional sparsity detection  and channel
	power) estimation): Over the entire long transmission frame, we assume that the statistical CSI between all 6DMA candidate position-rotation pairs and all users remains constant. In Stage I, we estimate the statistical CSI in terms of the directional sparsity matrix $\mathbf{Z}$ and the average channel power matrix \(\mathbf{P} \) (see Section IV-A for details). Specifically, the 6DMA surface moves over $\overline{M}\ll M$ different sampling position-rotation pairs to collect data for estimating the corresponding statistical CSI.
To reduce the CPU processing complexity and the required baseband transmission rate between the 6DMA surfaces and the CPU, we propose dividing the $\overline{M}$ position-rotation pairs into $B$ groups (see Fig. \ref{candidate}), with each group containing $M_{\mathrm{g}} = \overline{M}/B$ different position-rotation pairs\footnote{We assume that \( \overline{M} \) is divisible by \( B \) for convenience.}. Each 6DMA surface is assigned to a different group and it moves to  all $M_{\mathrm{g}}$
    different positions-rotations pairs within that group, while its LPU collects the uplink pilot signals received from the users (which have non-zero channels with it). Then, each LPU independently estimates the statistical CSI for its associated group of $M_{\mathrm{g}}$ 6DMA position-rotation pairs. This can be done in parallel for all LPUs. Subsequently, we can reconstruct the statistical CSI for all $M \gg \overline{M}$ possible position-rotation pairs in BS site region \(\mathcal{C}\), based on the previously estimated statistical CSI for the small number of \( \overline{M} \) position-rotation pairs.
    Note that the slow movement of 6DMA surfaces neither affects channel estimation accuracy nor interrupts continuous user communication. 
   
	\item  Stage II (6DMA position/rotation optimization and movement): After Stage I, all LPUs send their locally estimated candidate position/rotation statistical CSI to the CPU. With the global statistical CSI, the CPU of the 6DMA-BS then optimizes the positions and rotations of all $B$ 6DMA surfaces in this stage to maximize the ergodic sum rate (see Section V for details). Once the optimized positions and rotations of the 6DMA surfaces have been determined, the 6DMA surfaces are gradually moved over consecutive short communication blocks to the optimized positions and rotations. Note that, similar to Stage I, user communications continue uninterruptedly during this stage despite the movement of the 6DMA surfaces.

	\item  Stage III (User communication with improved data rate at optimized 6DMA positions/rotations and directional sparsity-aided instantaneous channel estimation): With all 6DMA surfaces located at their optimized positions and rotations obtained in Stage II, the users can communicate with the BS with an improved sum rate in Stage III, where the instantaneous channels between the users and each 6DMA surface located at its optimized  position and rotation can be estimated within a short communication block at the connected LPU in a distributed manner by leveraging the directional sparsity estimated in Stage I (see Section IV-B for details).
		
\end{itemize}

In the above proposed protocol, the 6DMA channel acquisition problem is decoupled into the statistical CSI estimation for all candidate positions/rotations and the subsequent  instantaneous CSI estimation for the optimized positions/rotations to exploit the 6DMA channel directional sparsity and reduce the pilot overhead. On the one hand, the channel comprising all candidate positions/rotations is high-dimensional, but its statistical properties do not change during the long transmission frame. Thus, the candidate position/rotation statistical CSI, needed for antenna position/rotation optimization, can be estimated much less frequently than the candidate position/rotation instantaneous CSI would have to be estimated, leading to a low average pilot overhead. On the other hand, for both candidate position/rotation statistical CSI estimation and optimized position/rotation instantaneous CSI estimation, the 6DMA channel directional sparsity can be exploited to further reduce the pilot overhead, as will be shown in the next section in detail.  

\section{Distributed Channel Estimation Algorithms}
In this section, we propose distributed 6DMA channel estimation schemes to acquire the statistical CSI for all candidate 6DMA positions and rotations, and the instantaneous CSI at the optimized 6DMA positions and rotations, as specified in Stages I and III in the proposed protocol (see Fig. \ref{timescale}), respectively.

\subsection{Joint Directional Sparsity Detection and Channel Power Estimation in Stage I}
The proposed distributed statistical CSI estimation algorithm is implemented in two steps. First, the 6DMA surface moves over $\overline{M} \ll M$ different sample position-rotation pairs to collect data for estimating the corresponding statistical CSI.
Second, based on the channel power estimated in Step 1 (see below), the multi-path average power and the DOA vector are determined for reconstructing the statistical CSI for all $M$ possible 6DMA positions and rotations.

\subsubsection*{Step 1: Covariance-Based Average Channel Power Estimation}
We first perform candidate position-rotation pair grouping. Specifically, to perform distributed channel estimation, we first divide the $\overline{M}$ position-rotation pairs into $B$ equal-size groups, with each group containing $M_{\mathrm{g}}=\frac{\overline{M}}{B}$ position-rotation pairs (see Fig. \ref{candidate}). The LPU connected to the $b$-th 6DMA surface is responsible for estimating the channels between all $K$ users and the subset of position-rotation pairs in group $b, b=1,2,\cdots,B$. In this way, all LPUs can estimate their corresponding channels in parallel, thus reducing the channel estimation complexity and pilot overhead in each group.

For the $b$-th position-rotation pair group, the corresponding 6DMA surface moves over a total of $M_{\mathrm{g}}$ different position-rotation pairs to collect channel measurement for estimation of the statistical CSI.
Specifically, to enable the directional sparsity detection and average channel power estimation in Stage I of the proposed protocol, we assume that each LPU collects in total $L$ pilot symbols from each user. It is 
assumed that the pilots from user $k$, denoted by $\mathbf{x}_k \in \mathbb{C}^{L\times 1}$, are generated following an independent and identically distributed (i.i.d.) complex Gaussian distribution with zero mean and unit variance. 	Let \( \mathcal{M}_b \) denote the set of indices corresponding to the position-rotation pairs assigned to the \( b \)-th LPU, where \( |\mathcal{M}_b| = M_{\mathrm{g}} \). Then, for any LPU, the received signal \( \mathbf{Y}_m \in \mathbb{C}^{L \times N} \), where \( m \in \mathcal{M}_b \), at the \( m \)-th candidate position-rotation pair can be expressed as
\begin{align}
	\label{aps}
	\mathbf{Y}_m&=\sum_{k=1}^K[\bar{\mathbf{Z}}]_{m,k}\mathbf{x}_k \mathbf{h}_{m, k}^T(\mathbf{q}_{m},\mathbf{u}_{m})+\mathbf{W}_m\nonumber\\
	&=\mathbf{X}\text{diag}([\bar{\mathbf{Z}}]_{m,:})\mathbf{H}_m^T+\mathbf{W}_m,
\end{align}
where $\bar{\mathbf{Z}}\in \mathbb{R}^{\overline{M} \times K}$ represents the directional sparsity indicator matrix for the $\overline{M}$ position-rotation pairs, 
$\mathbf{X}=[\mathbf{x}_1,\cdots,\mathbf{x}_K]\in \mathbb{C}^{L \times K}$ denotes the horizontal stack of all pilots from all users, and $\mathbf{W}_m \in \mathbb{C}^{L \times N}$ is the AWGN marix with i.i.d. entries following distribution $\mathcal{CN}(0,\sigma^2)$. 

{\bf{Remark 1}}:  Our goal is to estimate the 
average channel power matrix for $\overline{M}$ position-rotation pairs, denoted by $\bar{\mathbf{P}}\in \mathbb{R}^{\overline{M} \times K}$, and the directional sparsity indicator matrix $\bar{\mathbf{Z}}$ in Stage I, as they will be used as the statistical CSI for the subsequent statistical CSI reconstruction for all possible $M$ 6DMA positions and rotations. One possible approach to do this is to treat \eqref{aps} as a compressed sensing (CS) model with multiple measurement vectors and exploit the row sparsity of \(\tilde{\mathbf{H}}_m=\text{diag}([\bar{\mathbf{Z}}]_{m,:})\mathbf{H}_m^T \in \mathbb{C}^{K\times N}\). Once \(\tilde{\mathbf{H}}_m \) is recovered from \( \mathbf{Y}_m \) by applying CS-based algorithms, \([\bar{\mathbf{Z}}]_{m,:} \) can be determined from the rows of \( \tilde{\mathbf{H}}_m \). However, the computational complexity of this approach scales with \( N \) since \( \tilde{\mathbf{H}}_m \) has size \( K \times N \), which may be problematic for large \( N \). Moreover, the exact entries of \( \tilde{\mathbf{H}}_m\) are not required for the subsequent 6DMA position and rotation optimization (as it is based on the statistical CSI of the 6DMA channels), and estimating them  would require a high pilot overhead due to the large number of unknowns \cite{zhilin}. 

Therefore, instead of recovering the exact channel \( \tilde{\mathbf{H}}_m \), we estimate the  power state vector \(\boldsymbol{\eta}_m \) (from which \( [\bar{\mathbf{Z}}]_{m,k} \) can also be determined), which is defined as
\begin{align}\label{wq}
	&\boldsymbol{\eta}_m=\left[[\bar{\mathbf{P}}]_{m,1}[\bar{\mathbf{Z}}]_{m,1}, [\bar{\mathbf{P}}]_{m,2}[\bar{\mathbf{Z}}]_{m,2}, \cdots, [\bar{\mathbf{P}}]_{m,K}[\bar{\mathbf{Z}}]_{m,K}\right]^T \nonumber\\
	&~~~~~~~~~~~~~~~~~~~~~~~~~~~~~\in\mathbb{R}^{K\times 1}, m \in \mathcal{M}_b.
\end{align}

Since we assume that the channel vectors of different users are independent and follow a Gaussian distribution with zero mean, based on \eqref{wq}, each column of $\mathbf{Y}_m$, denoted as $[\mathbf{Y}_m]_{:,n}$, $1 \leq n \leq N$, can be treated as an independent sample following a multivariate complex Gaussian distribution based on \eqref{wq}, i.e., 
\begin{equation}\label{ml}
	[\mathbf{Y}_m]_{:,n}\sim \mathcal{CN}(\mathbf{0},\mathbf{X}\text{diag}(\boldsymbol{\eta}_m)\mathbf{X}^H+\sigma^2\mathbf{I}_L).
\end{equation}

In the following, we aim to jointly determine 6DMA directional sparsity matrix $\bar{\mathbf{Z}}$ and average channel power matrix $\bar{\mathbf{P}}$.
This involves estimating the power state vector $\boldsymbol{\eta}_m$ from the noisy observations $\mathbf{Y}_m$ by utilizing the known pilot matrix $\mathbf{X}$. In general, this estimation can be formulated as a maximum likelihood (ML) problem \cite{covar, zhilin}. 
Specifically, we define
$\boldsymbol{\Sigma}_m=\mathbf{X}\text{diag}(\boldsymbol{\eta}_m)\mathbf{X}^H+\sigma^2\mathbf{I}_L$.
Then, the likelihood function of $\mathbf{Y}_m$ given $\boldsymbol{\eta}_m$ can be represented as \footnote{The authors of \cite{zhilin} have shown that although \eqref{ui} is valid only for channels that are uncorrelated across the \(N\) antennas of the same 6DMA surface, the proposed covariance-based method still provides an accurate estimate of \(\boldsymbol{\eta}_m\) as long as the channel vectors are not highly correlated, such as under line-of-sight (LoS) conditions.}
\begin{subequations}
\begin{align}
	P(\mathbf{Y}_m|\boldsymbol{\eta}_m)
	&=\prod_{n=1}^{N}\frac{1}{\det(\pi\boldsymbol{\Sigma}_m)}\exp(-[\mathbf{Y}_{m}]_{:,n}^H\boldsymbol{\Sigma}_m^{-1}[\mathbf{Y}_m]_{:,n})\label{ui}\\
	&=\frac{1}{\det(\pi\boldsymbol{\Sigma}_m)^N}\exp(-\text{tr}(\boldsymbol{\Sigma}_m^{-1}\mathbf{Y}_m\mathbf{Y}_m^H)).\label{liki}
\end{align}
\end{subequations}

After performing normalization and simplification, we obtain
the following equation for the maximum likelihood estimator
for $\boldsymbol{\eta}_m$: 
\begin{align}\label{eqr}
	f(\boldsymbol{\eta}_m)=-\ln P(\mathbf{Y}_m|\boldsymbol{\eta}_m)=\ln\text{det}(\boldsymbol{\Sigma}_m)+\text{tr}(\boldsymbol{\Sigma}_m^{-1}\hat{\boldsymbol{\Sigma}}_{m}),
\end{align}
where $\hat{\boldsymbol{\Sigma}}_{m}=\frac{1}{N}\mathbf{Y}_m\mathbf{Y}_m^H$ denotes the sample covariance matrix of the received signal at the $m$-th candidate position/rotation pair averaged over all $N$ antennas of the 6DMA surface. Based on \eqref{eqr}, the ML estimation problem can be formulated as follows
\begin{align} \label{co}
	\arg \min_{\boldsymbol{\eta}_m\in\mathbb{R}_+}f(\boldsymbol{\eta}_m).
\end{align}
The solution to \eqref{co} depends on $\mathbf{Y}_m$ through sample covariance matrix $\frac{1}{N}\mathbf{Y}_m\mathbf{Y}_m^H$, whose size scales with $L$
instead of $N$. 

Next, we derive a closed-form expression for the coordinate-wise minimization of \( f(\boldsymbol{\eta}_m) \) in \eqref{eqr}. Let \( k \in \{1,2,\ldots,K\} \) be the index of the coordinate being considered and $\nu$ be the update step. We define \( f_k(\nu) = f(\boldsymbol{\eta}_m + \nu \mathbf{e}_k) \), where \( \mathbf{e}_k \in \mathbb{R}^{K} \) is the \( k \)-th canonical basis vector. Using the Sherman-Morrison rank-one update identity \cite{sher}, we obtain
\begin{align}\label{sdd}
	\!\!\!\!\!\!&&\!\!\!\!\!\!\left( \boldsymbol{\Sigma}_{m}+\nu\mathbf{x}_k\mathbf{x}_k^H\right )^{-1}=\boldsymbol{\Sigma}_{m}^{-1}-
	\frac{\nu\boldsymbol{\Sigma}_{m}^{-1}\mathbf{x}_k\mathbf{x}_k^H\boldsymbol{\Sigma}_{m}^{-1}}{1+\nu\mathbf{x}_k^H\boldsymbol{\Sigma}_{m}^{-1}\mathbf{x}_k}.
\end{align}
Applying the well-known determinant identity \cite{mat}, we have
\begin{align}\label{sdd1}
	\text{det}(\boldsymbol{\Sigma}_{m}+\nu\mathbf{x}_k\mathbf{x}_k^H)=(1+\nu\mathbf{x}_k^H\boldsymbol{\Sigma}_{m}^{-1}\mathbf{x}_k)\text{det}(\boldsymbol{\Sigma}_{m}).
\end{align}
Then, substituting \eqref{sdd} and \eqref{sdd1} into \eqref{eqr}, we can simplify $f_k(\nu)$ as follows
\begin{align}\label{fvu}
f_k(\nu)=\kappa_m +\ln(1+\nu\mathbf{x}_k^H\boldsymbol{\Sigma}_{m}^{-1}\mathbf{x}_k)-\frac{\mathbf{x}_k^H\boldsymbol{\Sigma}_{m}^{-1}\hat{\boldsymbol{\Sigma}}_{m}\boldsymbol{\Sigma}_{m}^{-1}\mathbf{x}_k}{1+\nu\mathbf{x}_k^H\boldsymbol{\Sigma}_{m}^{-1}\mathbf{x}_k}\nu,
\end{align}
where $\kappa_m =	\text{det}(\boldsymbol{\Sigma}_{m})+\text{tr}(\boldsymbol{\Sigma}_{m}^{-1}\hat{\boldsymbol{\Sigma}}_{m})$.
Taking the derivative of $f_k(\nu)$ w.r.t. $\nu$ leads to
\begin{align}\label{gsi}
	\bigtriangledown f_k(\nu)=\frac{\mathbf{x}_k^H\boldsymbol{\Sigma}_{m}^{-1}\mathbf{x}_k}
	{1+\nu\mathbf{x}_k^H\boldsymbol{\Sigma}_{m}^{-1}\mathbf{x}_k}-\frac{\mathbf{x}_k^H\boldsymbol{\Sigma}_{m}^{-1}\hat{\boldsymbol{\Sigma}}_{m}\boldsymbol{\Sigma}_{m}^{-1}\mathbf{x}_k}{(1+\nu\mathbf{x}_k^H\boldsymbol{\Sigma}_{m}^{-1}\mathbf{x}_k)^2}.
\end{align}
The solution of $\bigtriangledown f_k(\nu)=0$ is thus  given by 
\begin{align}\label{so}
\nu^{*}=\frac{\mathbf{x}_k^H\boldsymbol{\Sigma}_{m}^{-1}\hat{\boldsymbol{\Sigma}}_{m}\boldsymbol{\Sigma}_{m}^{-1}\mathbf{x}_k-\mathbf{x}_k^H\boldsymbol{\Sigma}_{m}^{-1}\mathbf{x}_k}{(\mathbf{x}_k^H\boldsymbol{\Sigma}_{m}^{-1}\mathbf{x}_k)^2}.
\end{align}
From \eqref{fvu}, we observe that \(f_k(\nu)\) is well-defined when \(\nu > \nu_0 := -\frac{1}{\mathbf{x}_k^H\boldsymbol{\Sigma}_{m}^{-1}\mathbf{x}_k}\). Consequently, it follows that \(f_k(\nu)\) is also well-defined at \(\nu = \nu^*\). Furthermore, from \eqref{fvu}, we observe that \(\lim_{\varepsilon \to 0^+} f_k(\nu_0 +\varepsilon) = \lim_{\nu \to +\infty} f_k(\nu) = +\infty\), which implies that \(\nu = \nu^*\) must be the global minimum of \(f_k(\nu)\) within \((\nu_0, \infty)\).
To ensure that \( [\boldsymbol{\eta}_m]_k \) remains positive after the update \( [\boldsymbol{\eta}_m]_k \to [\boldsymbol{\eta}_m]_k + \nu \), the optimal update step \(\nu\) is \( \max \left(\nu^*, -[\boldsymbol{\eta}_m]_k \right) \). 

\subsubsection*{Step 2: Multi-Path Average Power and DOA Vector Estimation}
Once the statistical CSI for a small number of \( \overline{M} \) position-rotation pairs, i.e., $\bar{\mathbf{P}}\in \mathbb{R}^{\overline{M} \times K}$, is estimated and transmitted to the CPU, the statistical CSI for all possible $M$ position-rotation pairs in the BS site region \(\mathcal{C}\) can be reconstructed at the CPU \cite{icc}, by formulating the reconstruction as a sparse signal recovery problem and solving it using compressed sensing techniques. Specifically, the elements of \(\bar{\mathbf{P}}\) can be rewritten as follows:
\begin{subequations}
	\begin{align}
		[\bar{\mathbf{P}}]_{m,k}
		&=\sum_{n=1}^N \mathbb{E}\left[\left|[\mathbf{h}_{m,k}(\mathbf{q}_{m},
		\mathbf{u}_{m})]_n\right|^2\right],\label{po}\\
		&=N \mathbb{E}\left[\left|[\mathbf{h}_{m,k}(\mathbf{q}_{m},
		\mathbf{u}_{m})]_n\right|^2\right], \forall n\in \mathcal{N}\label{po1}\\
		&=N g_{k}(\mathbf{u}_{m},\mathbf{f}_{k}) t_k, \label{mm}
	\end{align}
\end{subequations}
where $t_k=\mathbb{E}\left[\left| \sum_{\iota=1}^{\Gamma_{k}}\sqrt{\mu_{\iota, k}}e^{-j \bar{\varphi}_{\iota, k}} \right|^2\right]$ represents the multi-path average power from the $k$-th user to the 6DMA-BS, and $\bar{\varphi}_{\iota, k}={\varphi}_{\iota, k}+\frac{2\pi}{\lambda}
(\mathbf{f}_{\iota, k}-\mathbf{f}_{k})^T\mathbf{r}_{i,n}(\!\mathbf{q}_{m},
\mathbf{u}_{m})$, which is modeled as an independent and uniformly distributed random variable in \([0, 2\pi)\). Eq. \eqref{mm} holds because 
$[\mathbf{h}_{m,k}(\mathbf{q}_{m},
	\mathbf{u}_{m})]_n
	\approx \sqrt{g_{k}(\mathbf{u}_{m},\mathbf{f}_{k})}
	e^{-j\frac{2\pi}{\lambda}
		\mathbf{f}_{k}^T\mathbf{r}_{i,n}(\!\mathbf{q}_{m},
		\mathbf{u}_{m}\!)}\sum_{\iota=1}^{\Gamma_{k}}\sqrt{\mu_{\iota, k}}e^{-j \bar{\varphi}_{\iota, k}}$, 
where $g_{\iota,k}(\mathbf{u}_{m},\mathbf{f}_{\iota,k})$  is assumed to remain constant for all $\iota$, i.e., $g_{\iota,k}(\mathbf{u}_{m},\mathbf{f}_{\iota,k}) = g_{k}(\mathbf{u}_{m},\mathbf{f}_{k}), \forall \iota \in \{1, 2, \cdots, \Gamma_k\}$, with $\mathbf{f}_{k}$ denoting the unit DOA vector corresponding to the signal arriving at the BS from the center of the scattering cluster of user $k$.	

From \eqref{mm}, we know that the 6DMA channel power is a function of the multi-path average power \(t_k\) and the unit-length DOA vector \(\mathbf{f}_{k}\). Therefore, by determining \(\mathbf{f}_{k}\) and \(t_k\) based on the estimated $\bar{\mathbf{P}}$ and $\bar{\mathbf{Z}}$ in Step I, the average channel power between users and all $M$ possible 6DMA positions and rotations can be reconstructed at the CPU.
In particular, we leverage the knowledge of channel directional sparsity matrix $\bar{\mathbf{Z}}$ estimated in Step I to reduce the parameter estimation complexity in Step II.
Specifically, we construct the support vector \(\bar{\mathcal{I}}_k \in \mathbb{R}^{M_k}\) for \([\bar{\mathbf{Z}}]_{:,k}\), which includes the indices of its non-zero elements, with \(M_k < M\) denoting the number of non-zero elements in \([\bar{\mathbf{Z}}]_{:,k}\).
We define $\bar{\mathbf{p}}_k=[\bar{\mathbf{P}}]_{:,k}\in \mathbb{R}^{M}$ and $\mathbf{v}_k =[ g_{k}(\mathbf{u}_{1},\mathbf{f}_{k}),\cdots, g_{k}(\mathbf{u}_{M},\mathbf{f}_{k})]^T\in \mathbb{R}^{M}$. 
Then, let \(\bar{\mathbf{p}}_{k,\bar{\mathcal{I}}_k} \in \mathbb{R}^{M_k}\) and \(\mathbf{v}_{k,\bar{\mathcal{I}}_k} \in \mathbb{R}^{M_k}\) denote the new vectors formed by the non-zero elements of \(\bar{\mathbf{p}}_k\) and \(\mathbf{v}_k\) that correspond to the indices stored in \(\bar{\mathcal{I}}_k\). 
Consequently, we optimize 
$\mathbf{f}_{k}$ and $t_k$ by minimizing the reconstruction error:
\begin{subequations} \label{ac}
	\begin{align}
		~&~ \min_{ t_k, \mathbf{f}_{k} } \left\|\bar{\mathbf{p}}_{k,\bar{\mathcal{I}}_k} - N\mathbf{v}_{k,\bar{\mathcal{I}}_k} t_k  \right\|^2_2,\\
		~&~ \text{s.t.}~~~{t}_k\ge 0.
	\end{align}
\end{subequations}

To efficiently estimate $\mathbf{f}_{k}$ and $t_k$ by compressed sensing, we
approximate $\mathbf{f}_{k}$ by uniformly discretizing it into $G$ grid points with $G\geqslant 1$. Thus, we have $\mathbf{v}_{k,\bar{\mathcal{I}}_k} t_k  \approx \tilde{\mathbf{V}}_k \tilde{\mathbf{s}}_k$, where $\tilde{\mathbf{V}}_k\in\mathbb{R}^{M_k\times G}$
is an over-complete matrix and $\tilde{\mathbf{s}}_k\in\mathbb{R}^{G}$
is a sparse vector with one non-zero element corresponding to ${t}_k$.
Then, problem \eqref{ac} reduces to
\begin{subequations}
	\label{eg3}
	\begin{align}
		~&~ \arg \min_{\tilde{\mathbf{s}}_k} \| \bar{\mathbf{p}}_{k,\bar{\mathcal{I}}_k} - N\mathbf{\tilde{V}}_k \tilde{\mathbf{s}}_k   \|_2 \\
		~&~ \mathrm{s.t.} ~ \|\tilde{\mathbf{s}}_k\|_0 = 1, \\
		~&~~~~~~ \tilde{\mathbf{s}}_k \succeq \bf{0}.
	\end{align}
\end{subequations}
Problem \eqref{eg3} is a compressed sensing problem that can be solved using classical compressed sensing algorithms, such as non-negative orthogonal matching pursuit (OMP) \cite{let}, to estimate $\tilde{\mathbf{s}}_k$ and obtain the estimated $\hat{t}_k$. Subsequently, the estimated $\hat{\mathbf{f}}_k$, corresponding to the columns of $\tilde{\mathbf{V}}_k$ with non-zero coefficients in $\tilde{\mathbf{s}}_k$, can be obtained accordingly.

Finally, the estimate of \({\mathbf{P}} \in \mathbb{R}^{M \times K}\) for all possible positions and rotations can be expressed as
\begin{align}\label{esti}
	[\hat{\mathbf{P}}]_{m,k}
	=N g_{k}(\mathbf{u}_{m},\hat{\mathbf{f}}_{k}) \hat{t}_k.
\end{align}

We summarize the above  distributed joint directional sparsity detection and channel power estimation (JDC) algorithm in Algorithm 1, where in Line 6 a random selection strategy is adopted for index \(k\) to improve computational efficiency and ensure a sufficient number of updates across all indices over multiple iterations. The threshold $\epsilon$ for sparsity detection is empirically chosen based on system-specific observations and practical experience. Note that in Algorithm 1, all LPUs estimate the statistical CSI parameters in a parallel manner. This not only accelerates computation but also significantly reduces the computational overhead of each LPU, as the number of channels to be estimated for each LPU is reduced from \(\overline{M}\) to a smaller number \(M_{\mathrm{g}}\). 
\begin{algorithm}[t!]
	\caption{Distributed Joint Directional Sparsity Detection and Channel Power (JDC) Estimation}
	\label{alg1}
	\begin{algorithmic}[1]
		\STATE \textbf{Input}: $\mathbf{X}$, $\{\hat{\boldsymbol{\Sigma}}_{m}=\frac{1}{N}\mathbf{Y}_m\mathbf{Y}_m^H\}_{m=1}^{\overline{M}}$, and number of iterations $T$.
		\STATE \textbf{Initialization}: $\{\boldsymbol{\eta}_m=\mathbf{0}\}_{m=1}^{\overline{M}}$, $\bar{\mathbf{Z}}=\mathbf{0}$, $\hat{\mathbf{Z}}=\mathbf{0}$, $\{\boldsymbol{\Sigma}_m=\sigma^2\mathbf{I}_L\}_{m=1}^{\overline{M}}$. \\
		\STATE {For each LPU $b=1,2,\cdots, B$, do the following steps 4 to 12 in parallel:}
		\FOR{$t=1 : T$} 
		\FOR{each \( m \in \mathcal{M}_b \)} 
		\STATE {Select an index \( k \in \{1,2, \cdots, K\} \) corresponding to the \( k \)-th element of $\boldsymbol{\eta}_m$ randomly;}
		\STATE  Set $\nu^*=\max\left\{\frac{\mathbf{x}_k^H\boldsymbol{\Sigma}_{m}^{-1}\hat{\boldsymbol{\Sigma}}_{m}\boldsymbol{\Sigma}_{m}^{-1}\mathbf{x}_k-\mathbf{x}_k^H\boldsymbol{\Sigma}_{m}^{-1}\mathbf{x}_k}{(\mathbf{x}_k^H\boldsymbol{\Sigma}_{m}^{-1}\mathbf{x}_k)^2},-[\boldsymbol{\eta}_m]_k\right\}$;
		\STATE  Update $[\boldsymbol{\eta}_m]_k=[\boldsymbol{\eta}_m]_k+\nu^*$;
		\STATE  $\boldsymbol{\Sigma}_m=\boldsymbol{\Sigma}_m+\nu^*\mathbf{x}_k\mathbf{x}_k^H$;
	\STATE Set directional sparsity \( [\bar{\mathbf{Z}}]_{m,k} = 1 \) if \( [\boldsymbol{\eta}_m]_k \) exceeds a given threshold $ \epsilon >0$;
		\ENDFOR
		\ENDFOR
		\STATE Obtain $\bar{\mathbf{P}}\in \mathbb{R}^{\overline{M} \times K}$ and  $\bar{\mathbf{Z}}\in \mathbb{R}^{\overline{M} \times K}$;
		\STATE Reconstruct the channel power  for  all possible 6DMA positions and rotations,  $\hat{\mathbf{P}}\in \mathbb{R}^{M \times K}$,  according to \eqref{esti};
		\STATE Reconstruct the directional sparsity for all possible 6DMA positions and rotations by setting \( [\hat{\mathbf{Z}}]_{m,k} = 1 \) if \( [\hat{\mathbf{P}}]_{m,k} \) exceeds a given threshold $ {\epsilon} >0$;
		\STATE \textbf{Output}: $\hat{\mathbf{P}}\in \mathbb{R}^{M \times K}$ and $\hat{\mathbf{Z}}\in \mathbb{R}^{M \times K}$.
	\end{algorithmic}
\end{algorithm}

\subsection{Directional Sparsity Aided Instantaneous Channel Estimation in Stage III}
With the estimated directional sparsity from Stage I (see the preceding subsection) and the optimized 6DMA positions/rotations from Stage II (see Section V for details), in this subsection, we propose a directional sparsity-aided least-square (LS) algorithm for each LPU to estimate the instantaneous CSI from the users to its respective 6DMA surface at the optimized position/rotation in Stage III. Note that all LPUs can estimate the instantaneous channels between the users and their respective 6DMA surfaces simultaneously in parallel. Consequently, for any LPU $b$, we express the received signal $\mathbf{Y}_{i_b}\in \mathbb{C}^{L\times N}$ in terms of the transmitted pilot (measurement) matrix $\mathbf{X}$ as
\begin{align}
	\label{apse}
	\mathbf{Y}_{i_b}=\mathbf{X}\mathbf{H}_{i_b}^T+
	\mathbf{W}_{i_b}.
\end{align}
By vectorizing the matrix in \eqref{apse}, we have
\begin{align}
\mathbf{y}_{i_b}=\mathbf{A}\mathbf{h}_{i_b}
+\mathbf{w}_{i_b}, \label{uyw}
\end{align}
where $\mathbf{y}_{i_b}=\text{vec}(\mathbf{Y}_{i_b})$, $\mathbf{w}_{i_b}=\text{vec}(\mathbf{W}_{i_b})$, $\mathbf{h}_{i_b}=\text{vec}({\mathbf{H}}_{i_b}^T)\in\mathbb{C}^{NK\times 1}$,
and $\mathbf{A}=\mathbf{I}_{K}\otimes{\mathbf{X}}\in\mathbb{C}
^{NL\times KN}$. With the known $\mathbf{A}$, we aim to estimate $\mathbf{h}_{i_b}$ from $\mathbf{y}_{i_b}, b\in \mathcal{B}$.

Next, we leverage the knowledge of channel directional sparsity indicator matrix $\hat{\mathbf{Z}}$ estimated in the previous Stage I to improve the instantaneous channel estimation accuracy in Stage III. Specifically, we construct the support vector of $\mathbf{h}_{i_b}$, i.e.,  \(\hat{\mathcal{I}}_b \), 
based on the estimated directional sparsity indicator matrix $\hat{\mathbf{Z}}$ in Algorithm 1 as follows:
\begin{equation}\label{above}
	\hat{\mathcal{I}}_b=\mathrm{support}\left([\hat{\mathbf{Z}}]_{i_b,:}^T\otimes  \mathbf{1}_{N \times 1}\right)\in \mathbb{C}^{\bar{K}\times 1},
\end{equation}
where \(\hat{\mathcal{I}}_b\) identifies the indices of non-zero elements in \([\hat{\mathbf{Z}}]_{i_b,:}^T \otimes \mathbf{1}_{N \times 1}\), and \(\bar{K} < NK\) represents the number of these non-zero elements.

Then, let $\mathbf{A}_{\hat{\mathcal{I}}_b}$ denote the matrix composed of the corresponding columns of $\hat{\mathcal{I}}_b$ in matrix $\mathbf{A}$, and let $\mathbf{h}_{i_b,\hat{\mathcal{I}}_b}$ denote the vector composed of the
corresponding rows of $\hat{\mathcal{I}}_b$ in vector ${\mathbf{h}}_{i_b}$. Consequently, the signal in \eqref{uyw} can be rewritten as
\begin{align}
\mathbf{y}_{i_b}=\mathbf{A}_{\hat{\mathcal{I}}_b}
\mathbf{h}_{{i_b},\hat{\mathcal{I}}_b}
+\mathbf{w}_{i_b}. \label{yyc}
\end{align}
Last, we obtain the support-restricted estimates at the $b$-th ($b\in \mathcal{B}$) LPU using the following LS-based rule: 
\begin{align}
& \mathbf{h}_{i_b,\hat{\mathcal{I}}_b}\leftarrow\arg \min_{\mathbf{h}} \|\mathbf{A}_{\hat{\mathcal{I}}_b}\mathbf{h}-
\mathbf{y}_{i_b}\|_2^2,\nonumber\\
& \text{and}~~\mathbf{h}_{i_b,\hat{\mathcal{I}}_b^{\mathrm{c}}}\leftarrow 0, \label{ops}
\end{align}
where $
\hat{\mathcal{I}}_b \cup \hat{\mathcal{I}}_b^{\mathrm{c}} = \{1,2,\cdots,NK\}$ and $ \hat{\mathcal{I}}_b \cap \hat{\mathcal{I}}_b^{\mathrm{c}} = \emptyset
$. The resulting closed-form support-restricted estimates is given by \( \mathbf{h}_{i_b,\hat{\mathcal{I}}_b} = \mathbf{A}_{\hat{\mathcal{I}}_b}^\dagger \mathbf{y}_{i_b} \) with \( \mathbf{A}_{\hat{\mathcal{I}}_b}^\dagger = (\mathbf{A}_{\hat{\mathcal{I}}_b}^H \mathbf{A}_{\hat{\mathcal{I}}_b})^{-1} \mathbf{A}_{\hat{\mathcal{I}}_b}^H \). The full estimated vector \( \hat{\mathbf{h}}_{i_b} \) is then reconstructed as \( \hat{\mathbf{h}}_{i_b} = [\mathbf{h}_{i_b,\hat{\mathcal{I}}_b}^T, \mathbf{h}_{i_b,\hat{\mathcal{I}}_b^{\mathrm{c}}}^T]^T \). 
Note that since matrix \(\mathbf{A}_{\hat{\mathcal{I}}_b}\) has full column rank, \( \bar{K}\leq NL \) ensures that the LS solution in \eqref{ops} is well-defined and unique.

\subsection{Complexity Analysis}
\begin{table}[!t]
\small
	\caption{{Computational complexity Comparison of Different Algorithms.}}
	\label{Table1}
	\centering
	\begin{tabular}{|c|c|}
		\hline
		\makecell{Channel Estimation Algorithm} & \makecell{Complexity Order}
\\
			
\hline
		Distributed JDC (proposed)   & $\mathcal{O}(L^2KM_{\mathrm{g}}+M_{\mathrm{a}}G)$ \\
\hline
Distributed  AMP & $\mathcal{O}(LKNM_{\mathrm{g}}+M_{\mathrm{a}}G)$ \\
\hline
Centralized  BOMP & $\mathcal{O}(\overline{M}NKLL_h+M_{\mathrm{a}}G)$ \\
\hline
	\end{tabular}
\end{table}
In this subsection, we compare the computational complexity
of the proposed distributed JDC algorithm with those of two CS-based algorithms adopted as benchmarks, i.e., the distributed approximate message passing (AMP) algorithm \cite{vamp} and the centralized block OMP (BOMP) algorithm \cite{bomp}, for estimation of the statistical 6DMA channels. Both benchmark algorithms are two-step algorithms. For the distributed AMP algorithm, each LPU in Step I estimates the row-sparse channel \( \tilde{\mathbf{H}}_m \) in parallel using AMP. Once \(\tilde{\mathbf{H}}_m\) is recovered, the corresponding channel power can be determined based on it.
For the centralized BOMP algorithm, the CPU first collects signals from all LPUs, forming \(\mathbf{H}_{\mathrm{c}} = [\tilde{\mathbf{H}}_{1}^T, \cdots, \tilde{\mathbf{H}}_{M_\mathrm{g}}^T]^T \in \mathbb{C}^{\overline{M}K \times N}\). In Step I, the CPU then estimates the row-sparse channel \(\mathbf{H}_{\mathrm{c}}\) using BOMP. Once \(\mathbf{H}_{\mathrm{c}}\) is recovered, the corresponding channel power can be determined based on it.
Step II for both the distributed AMP and centralized BOMP algorithms remains the same as in the proposed algorithm.

We assume that the computational complexity is dominated by the required number of real-valued
multiplications. The computational complexity of the centralized BOMP algorithm is $\mathcal{O}(\overline{M}NKLL_h+M_{\mathrm{a}}G)$ \cite{oomp}, where $L_h$ denotes the number of non-zero blocks in $\mathbf{h}=\mathrm{vec}(\mathbf{H}^T)\in \mathbb{C}^{\bar{M}NK\times 1}$ and $M_{\mathrm{a}}=\max\{M_1, M_2, \cdots, M_K\}$. Besides, the computational complexity of the distributed AMP algorithm is $\mathcal{O}(LKNM_{\mathrm{g}}+M_{\mathrm{a}}G)$ \cite{vamp}.
In contrast, the computational complexity of our proposed  distributed JDC algorithm can be shown to be  $\mathcal{O}(L^2KM_{\mathrm{g}}+M_{\mathrm{a}}G)$. In Table I, we summarize the computational complexities of the considered schemes. We observe that the computational complexity of the proposed distributed JDC algorithm, as well as that of the two baseline algorithms, is proportional to the number of users $K$. However, the proposed distributed JDC algorithm exhibits significantly lower computational complexity compared to the two benchmarks for two main reasons. First, the proposed algorithm’s complexity does not depend on the number of antennas $N$ of each 6DMA surface, whereas both the conventional distributed AMP algorithm and centralized BOMP algorithm scale linearly with $N$.
	Second, while the complexities of the proposed distributed JDC algorithm and the conventional distributed AMP algorithm scale linearly with the number of position-rotation pairs per group, $M_g$, the complexity of the conventional centralized BOMP algorithm scales linearly with the total number of sample position-rotation pairs, $\overline{M} = B M_g$, which thus makes the proposed algorithm more computationally efficient as $\overline{M}$ is generally significantly larger than $M_g$.

\section{Channel Power-Based 6DMA Position and Rotation Optimization in Stage II}
After obtaining the estimate of the average channel power matrix $\mathbf{P}$ from the distributed LPUs in Stage I, the CPU optimizes the positions/rotations of all $B$ 6DMA surfaces (i.e., the position/rotation selection vector $\mathbf{s}$) in Stage II to maximize the ergodic sum rate upper bound $\bar{C}(\mathbf{s})$, subject to the practical movement constraints. Accordingly, the optimization problem is formulated as follows:
\begin{subequations}
\label{MG3}
\begin{align}
\text{(P1)}~~&~\mathop{\max}\limits_{\mathbf{s}}~~
\sum_{k=1}^K\log_2 \left( 1+\frac{p}{\sigma^2}\sum_{m=1}^{M}[\mathbf{s}]_m[\mathbf{P}]_{m,k}\right)\\
\text {s.t.}~&~\mathbf{1}^T\mathbf{s}=B, \label{M1}\\
~&~ \mathbf{e}_i^T(\mathbf{s})
\mathbf{D}\mathbf{e}_j(\mathbf{s})\geq d_{\min},~i\neq j,\forall i,j\in\mathcal{B},\label{M2}\\
~&~ [\mathbf{s}]_i\in\{0,1\}, i=1,2,\cdots,M, \label{M3}
\end{align}
\end{subequations}
where \( \mathbf{e}_i(\mathbf{s}) \in \mathbb{R}^{M \times 1} \) is a vector  with a single one at the index of the \( i \)-th non-zero entry of $\mathbf{s}$ and zeros elsewhere. For example, if \( \mathbf{s} = [1, 0, 0, 1, 0]^T\), then \(\mathbf{e}_1 = [1, 0, 0, 0, 0]^T\) and \(\mathbf{e}_2 = [0, 0, 0, 1, 0]^T\).
We define matrix
$\mathbf{D}\in\mathbb{C}^{M \times M}$, whose element $[\mathbf{D}]_{m,m'}=\|\mathbf{q}_m-\mathbf{q}_{m'}\|_2$ in the $m$-th row and $m'$-th column denotes the distance between the $m$-th discrete position and the $m'$-th discrete position.
As discussed in Section II-A,  constraint \eqref{M2} prevents overlapping and coupling between different 6DMA surfaces. Constraints \eqref{M1} and \eqref{M3} ensure that no candidate position-rotation pair is selected by more than one 6DMA surface. Note that for simplicity, in this paper, we assume that discrete candidate position-rotation pairs are generated uniformly on a sphere inside $\mathcal{C}$ (see Section VI for details). Thus, the rotation constraints in \eqref{rcc} and \eqref{dd} are naturally satisfied.

We note that (P1) is a non-convex integer programming problem that becomes progressively more challenging to solve optimally as the value of $M$ increases. In order to tackle this issue, we employ linear programming relaxation by relaxing the binary variables to take on continuous values within the range of [0,1] \cite{6dma_dis}. Then, inspired by the efficiency and effectiveness of particle swarm optimization (PSO) algorithms \cite{con}, we introduce a channel power-based PSO technique for optimizing the 6DMA positions and rotations. The proposed algorithm is described in Algorithm 2, which systematically updates many particles, each characterized by their position $\mathbf{s}\in\mathbb{R}^{M\times 1}$ and velocity $\boldsymbol{\xi}\in\mathbb{R}^{M\times 1}$. 

We aim to minimize $\bar{C}(\mathbf{s})$, which serves as the fitness function. However, due to the constraints in (P1), the fitness function must be adjusted \cite{con}. To satisfy constraints \eqref{M1}--\eqref{M3}, we incorporate an adaptive penalty factor, resulting in the following fitness function:
\begin{align}\label{ada}
	\mathcal{F}(\mathbf{s})=
	\bar{C}(\mathbf{s})+
	\tau\mathcal{Q}(\mathbf{s}),
\end{align}
where $\tau$ denotes the positive penalty parameter, and $\mathcal{Q}(\mathbf{s})$ is the penalty factor for the infeasible particles that violate the minimum
distance constraint \eqref{M2} and cardinality constraint \eqref{M1}, which is defined as
\begin{align}\label{pr}
	\mathcal{Q}(\mathbf{s}) 
	&=\left|\{(i,j)|
	\mathbf{e}_i(\tilde{\mathbf{s}})^T
	\mathbf{D}\mathbf{e}_j(\tilde{\mathbf{s}})< d_{\min}, 1 \leq i<j \leq B\}\right|_c \nonumber\\
	&+\left| B - \mathbf{1}^T \tilde{\mathbf{s}} \right|,
\end{align}
where $\tilde{\mathbf{s}} = \mathrm{round}(\mathbf{s})$.

In iteration $t$ of Algorithm 2, the position and velocity of the $\iota$-th particle are denoted by $\mathbf{s}_{\iota}^{(t)}$ and $\boldsymbol{\xi}_{\iota}^{(t)}$, respectively. As detailed in Algorithm 2, during initialization, we randomly set the positions and velocities of $I$ particles as follows:
\begin{align}
\mathcal{R}^{(0)}&=\{\mathbf{s}_1^{(0)},\mathbf{s}_2^{(0)},\cdots,
\mathbf{s}_I^{(0)}\},\label{rr0}\\
\mathcal{V}^{(0)}&=\{\boldsymbol{\xi}_1^{(0)},\boldsymbol{\xi}_2^{(0)},\cdots,
\boldsymbol{\xi}_I^{(0)}\}.\label{vv0}
\end{align}
Let $\mathbf{s}_{\iota,pbest}$ denote the best position of the $\iota$-th particle, and the $\mathbf{s}_{gbest}$ represent the global best position. In the initial stage, we set $\mathbf{s}_{\iota,pbest}=\mathbf{s}_{\iota}^{(0)}$, $1 \leq \iota\leq I $, and 
\begin{align}\label{ghl}
	\mathbf{s}_{gbest}=\arg \min\limits_{\mathbf{s}_{\iota}^{(0)}} \left\{ \mathcal{F}(\mathbf{s}_{1}^{(0)}),\mathcal{F}(\mathbf{s}_{2}^{(0)}),\cdots,
	\mathcal{F}(\mathbf{s}_{I}^{(0)})\right\}.
\end{align}
In each iteration $t$, the velocity of each particle $\iota$ is updated as \cite{con}
\begin{align}\label{gb}
\boldsymbol{\xi}_\iota^{(t+1)}&=\kappa\boldsymbol{\xi}_\iota^{(t)}+
c_1\tau_1(\mathbf{s}_{\iota,pbest}
-\mathbf{s}_\iota^{(t)})+c_2\tau_2,
\end{align}
where $c_1$ and $c_2$ are the individual and global learning factors, $\tau_1$ and $\tau_2$ are random parameters uniformly distributed in $[0, 1]$, and $\kappa$ is the inertia weight \cite{con}. In each iteration $t$, the position of each particle $\iota$ is updated as follows:
\begin{align}
\mathbf{s}_\iota^{(t+1)}&=\mathbf{s}_\iota^{(t)}+
\boldsymbol{\xi}_\iota^{(t)}.\label{vb}
\end{align}

Next, we evaluate each particle’s fitness using \eqref{ada} and compare it with the fitness values of the local and global best positions, as shown in Lines 10-15 of Algorithm 2. The global best fitness value is non-increasing, i.e., $\mathcal{F}(\mathbf{s}_{gbest}^{t+1}) \leq \mathcal{F}(\mathbf{s}_{gbest}^{t})$, which ensures convergence. The complexity of the proposed PSO algorithm is given by $\mathcal{O}(IT_{\mathrm{PSO}})$, where $T_{\mathrm{PSO}}$ is the maximum number of iterations of Algorithm 2.
\begin{algorithm}[t!]
	\caption{Channel Power-Based 6DMA PSO Algorithm for Solving (P1).}
	\label{alg3}
	\begin{algorithmic}[1]
		\STATE \textbf{Input}: $B$, $N$, $I$, $K$, $T_{\mathrm{PSO}}$, and channel power estimate $\hat{\mathbf{P}}$.  \\
		\STATE \textbf{Output}: Position/rotation selection vector $\mathbf{s}$.  \\
		\STATE Initialize particles with positions $\mathcal{R}^{(0)}$ and velocities
		$\mathcal{V}^{(0)}$ according to \eqref{rr0} and \eqref{vv0}, respectively.
		\STATE Obtain the local best position $\mathbf{s}_{\iota,pbest}=\mathbf{s}_{\iota}^{(0)}$ for $1 \leq \iota\leq I $ and the global best position according to \eqref{ghl};
		\FOR{$t = 1$ to $T_{\mathrm{PSO}}$}
		\FOR{$\iota = 1$ to $I$}
		\STATE Update the velocity of particle $\iota$ according
		to \eqref{gb};
		\STATE Update the position of particle $\iota$ according
		to \eqref{vb};
		\STATE Evaluate the fitness value of particle $\iota$, i.e.,
		$\mathcal{F}(\mathbf{s}_{\iota}^{(t)})$
		and update it according to \eqref{ada};
		\IF {$\mathcal{F}(\mathbf{s}_{\iota}^{(t)})>\mathcal{F}(\mathbf{s}_{\iota,pbest})$}
		\STATE Update $\mathbf{s}_{\iota,pbest}\leftarrow \mathbf{s}^{(t)}$;
		\ENDIF
		\IF {$\mathcal{F}(\mathbf{s}_{\iota}^{(t)})>\mathcal{F}(\mathbf{s}_{gbest})$}
		\STATE Update $\mathbf{s}_{gbest}\leftarrow \mathbf{s}^{(t)}$;
		\ENDIF
		\ENDFOR
		\ENDFOR
		\STATE Obtain the optimized rotation and position vector $\mathbf{s}=\mathbf{s}_{gbest}$;
		\STATE Return $\mathbf{s}=\text{round}(\mathbf{s})$.
	\end{algorithmic}
\end{algorithm}

\section{Simulation Results}
\begin{figure}[t!]
	\centering
	\setlength{\abovecaptionskip}{0.cm}
	\includegraphics[width=2.5in]{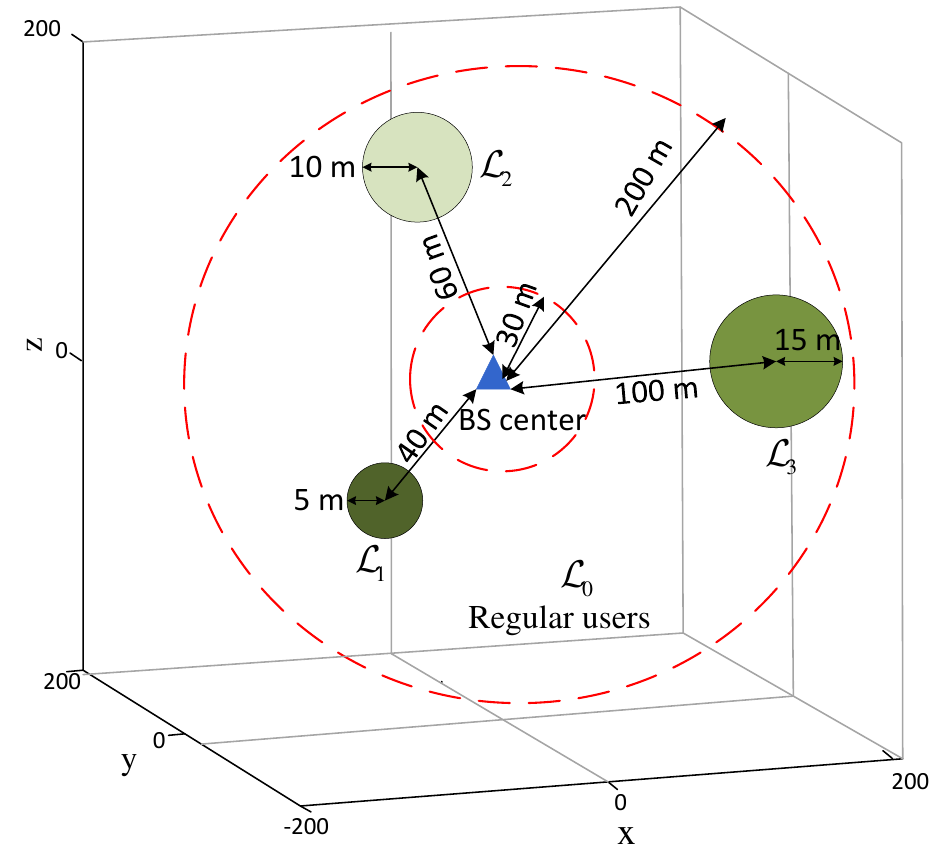}
	\caption{Simulation setup for 6DMA system. The darker the color of a hotspot area, the higher the user density in it.}
	\label{step}
\end{figure}

In this section, numerical results are provided to validate
the performance of our proposed 6DMA channel estimation algorithms as well as the statistical channel power based 6DMA position/rotation optimization.
Unless stated otherwise, we set \( N = 4 \), meaning each 6DMA surface is equipped with a \( 2 \times 2 \) UPA with antenna elements spaced \(d= \lambda/2 \). The number of 6DMA surfaces is \( B = 16 \), and the number of candidate position-rotation pairs is $M=256$.
The carrier frequency is \( 2.4 \) GHz and the wavelength is \( \lambda = 0.125 \) m. We set $\overline{M}=32$ and the 6DMA-BS site space \(\mathcal{C}\) as a cube with a side length of \(A = 1\) m.
We set the number of multi-path components per user as  $\Gamma_{k}=100, \forall k$ \footnote{Note that the proposed directional sparsity property, as well as the channel estimation and position/rotation optimization algorithms, remain also applicable and effective under LoS channel conditions.}. The multi-path channels for each user are generated by firstly randomly generating the user's location in the BS's coverage area (assumed to be a spherical annulus region $\mathcal{L}$ with radial distances between 30 m and 200 m from the CPU center) and then generating random scatterers uniformly within a circle centered at the user's location and with the radius equal to 3 m. As shown in Fig. \ref{step}, this coverage region $\mathcal{L}$ consists of three distinct hotspot sub-regions, $\mathcal{L}_v$ for $v=1,2,3$, and a regular user sub-region $\mathcal{L}_0$, such that $\mathcal{L}=\mathcal{L}_0\cup(\cup_{v=1}^3\mathcal{L}_v)$. The hotspot sub-regions $\mathcal{L}_1$, $\mathcal{L}_2$, and $\mathcal{L}_3$ are 3D spheres centered at 100 m, 60 m, and 40 m from the CPU, with radii of 15 m, 10 m, and 5 m, respectively. Users are distributed in these areas according to a homogeneous Poisson point process \cite{free6DMA}, with the proportion of regular users given by $ \varpi =\frac{\text{number of regular users}}{\text{total number of users}}=0.3$. The effective antenna gain $A(\tilde{\theta}_{b,\iota,k}, \tilde{\phi}_{b,\iota,k})$ in \eqref{gm} is set as the half-space directive antenna pattern \cite{3gpp, yumeng,shao20246d}.
 
\begin{figure}[t!]
	\centering
	\setlength{\abovecaptionskip}{0.cm}
	\includegraphics[width=2.5in]{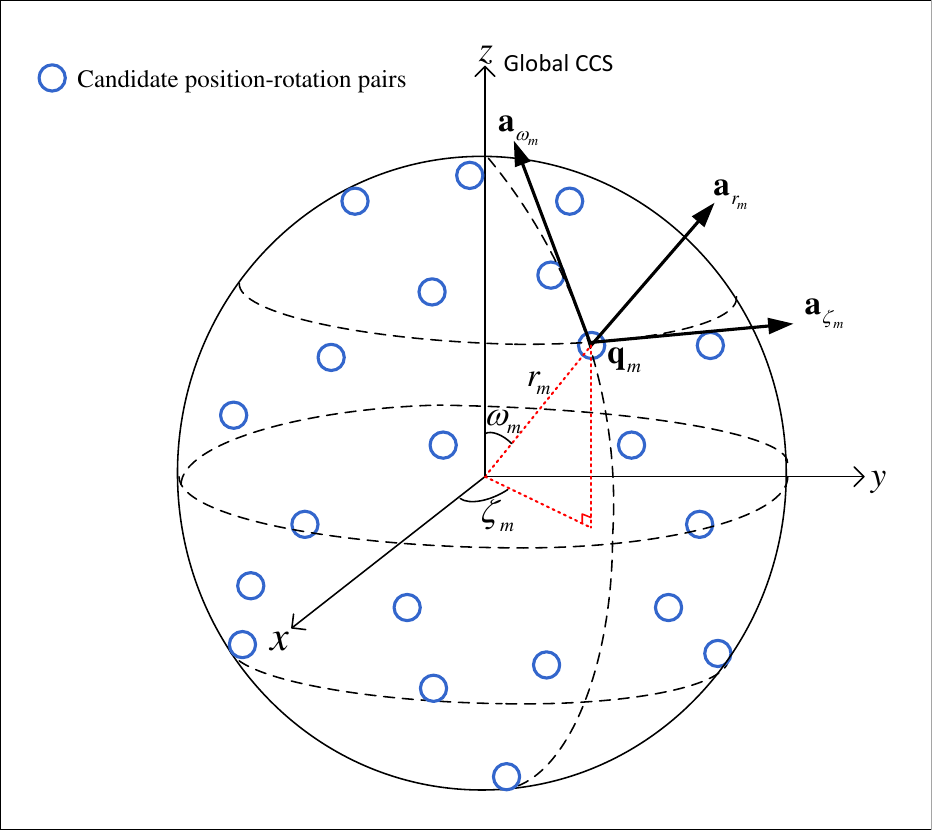}
	\caption{Candidate position-rotation pairs for 6DMA surfaces.}
	\label{sphere}
\end{figure}

For simplicity, we assume that discrete candidate position-rotation pairs are generated uniformly on a sphere. As shown in Fig. \ref{sphere}, we start by creating $M>B$ candidate positions evenly spread out on the biggest spherical surface that fits into the 6DMA-BS site space $\mathcal{C}$. We change the Cartesian coordinates of each possible position $\mathbf{q}_m, m\in \mathcal{M}$, to spherical coordinates $(r_m,w_m,\zeta_m)$ with $r_m$, $w_m$, and $\zeta_m$ being the position's radius, polar angle, and azimuth angle, respectively. We can then determine a unique rotation angle for each position based on the basis vectors of the spherical coordinates, denoted by $(\mathbf{a}_{r_m}, \mathbf{a}_{w_m}, \mathbf{a}_{\zeta_m})$. For further details, please refer to \cite{shao20246d}.

Now, we introduce the performance metrics for directional sparsity estimation, average channel power estimation, and instantaneous channel estimation. For directional sparsity detection, we adopt the detection error rate as performance metric. The detection error rate is defined as the sum of the missed detection probability, which is the probability that a user has a non-zero channel with a candidate position/rotation pair but is declared to have zero channel gain with it, and the false-alarm probability, which is the probability that a user has zero channel gain with a 6DMA position/ration pair but is declared to have a non-zero channel with it.

The performance of the average channel power estimation in Stage I
is evaluated by the normalized mean square error (NMSE),
which is defined as 
$\text{NMSE}_{\text{p}}=\mathbb{E}\left(\frac{\|
		\mathbf{P}-
		\hat{\mathbf{P}}\|_F^2}{\|\mathbf{P}\|_F^2}\right)$,
where $\hat{\mathbf{P}}$ denotes the estimated value of channel power matrix $\mathbf{P}$. 
Similarly, the performance of the instantaneous CSI estimation in Stage III
is evaluated by another NMSE,
which is defined as $
	\text{NMSE}_{\text{c}}=\mathbb{E}\left(\frac{\|\mathbf{h}-
		\hat{\mathbf{h}}\|_2^2}{\|\mathbf{h}\|_2^2}\right)$,
where $\hat{\mathbf{h}}$ denotes the estimated value of instantaneous channel vector $\mathbf{h}$. 

\begin{figure}[t!]
	\centering
	\setlength{\abovecaptionskip}{0.cm}
	\includegraphics[width=3.2in]{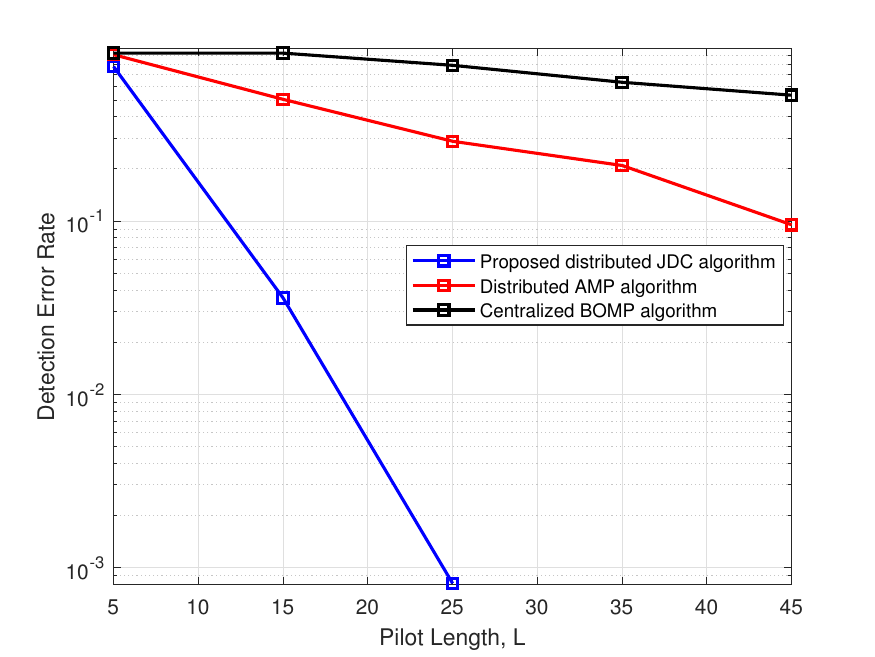}
	\caption{Detection  error rate versus pilot length with $K=50$ users.}
	\label{AER}
\end{figure}

We first conduct simulations to validate the effectiveness of the proposed JDC algorithm (Algorithm 1) for directional sparsity estimation. We also compare the proposed JDC algorithm with two baseline schemes (as discussed in Section IV-C): the distributed AMP algorithm \cite{vamp} and the centralized BOMP algorithm \cite{bomp}. Fig. \ref{AER} shows the estimation performance versus the length of the pilot sequence, $L$, under an average received signal-to-noise ratio (SNR) of 25 dB at the BS. From this figure, we can observe that the detection error rates of all the considered algorithms decrease as the pilot length increases. Moreover, the proposed JDC algorithm requires considerably fewer pilots for achieving the desired accuracy of directional sparsity estimation compared with the two benchmark schemes. Thus, the proposed JDC algorithm is more suitable for small $L$, for which AMP and BOMP may not work well. Furthermore, the proposed distributed JDC algorithm requires only the sample covariance of the channel observations; thus, it is quite robust w.r.t. variations in the statistics of the user channel vectors compared to the AMP algorithm, which depends heavily on the channel statistics for deriving the denoiser.

\begin{figure}[t!]
	\centering
	\setlength{\abovecaptionskip}{0.cm}
	\includegraphics[width=3.2in]{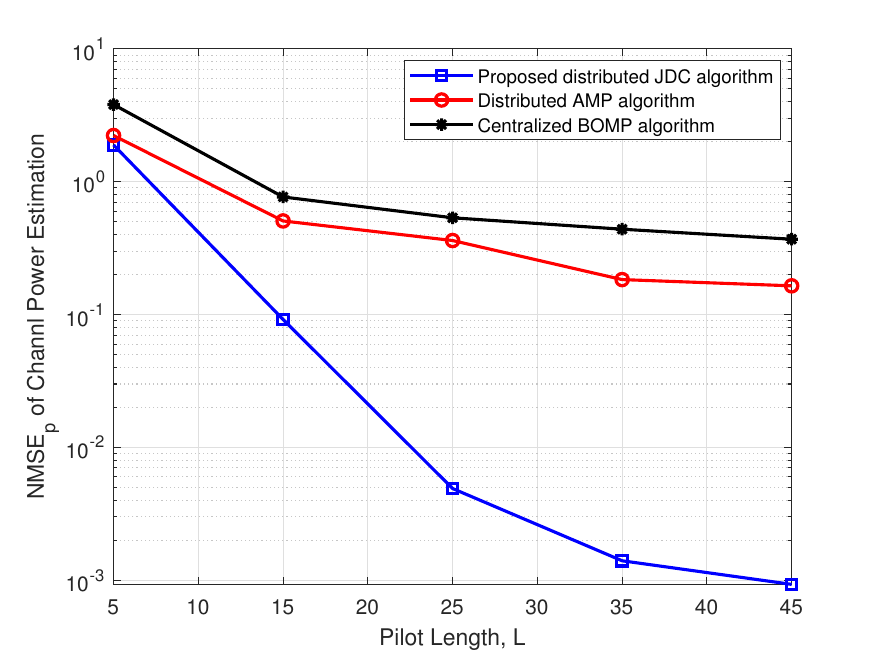}
	\caption{NMSE of channel power estimation versus pilot length for $K=50$ users and an average received SNR of 25 dB.}
	\label{NMSE_power}
\end{figure}

In Fig. \ref{NMSE_power}, we compare the proposed JDC algorithm with AMP and BOMP for average channel power estimation as a function of the pilot length $L$. We observe that, for the proposed JDC algorithm, increasing $L$ decreases the NMSE for channel power estimation  much faster  compared to the two benchmark schemes.
We can also see that, compared with the AMP and BOMP benchmark schemes, the proposed JDC algorithm reduces the length of the pilot sequence required to achieve a desired channel reconstruction accuracy. This improvement is due to the fact that the AMP and BOMP benchmark schemes have to estimate a much larger number of unknowns to recover the exact entries of \( \tilde{\mathbf{H}}_m \) and \( \mathbf{H}_{\mathrm{c}} \), respectively.

\begin{figure}[t!]
	\centering
	\setlength{\abovecaptionskip}{0.cm}
	\includegraphics[width=3.2in]{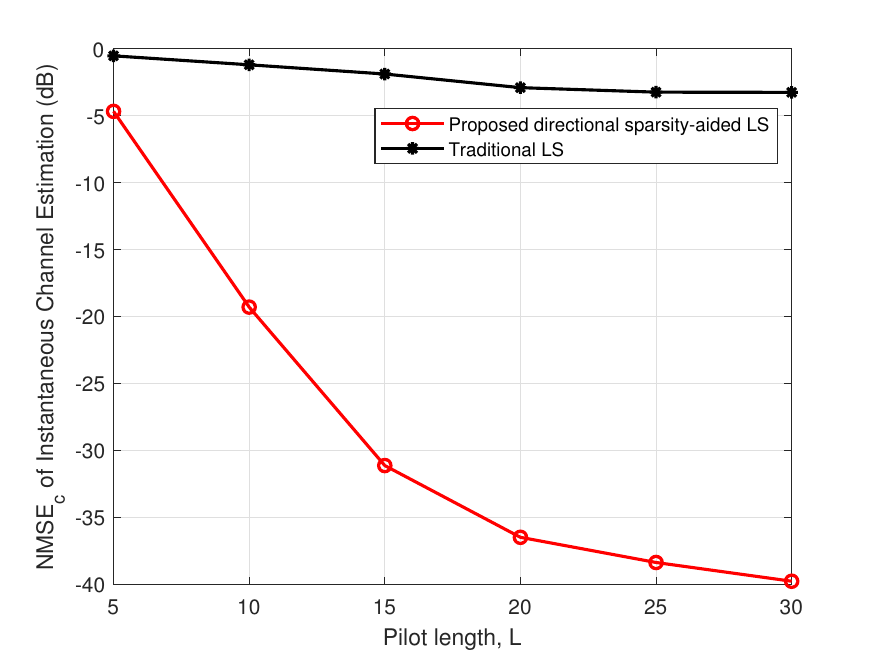}
	\caption{NMSE of instantaneous channel estimation versus  pilot length for $K=30$ users and an average received SNR of 25 dB.}
	\label{actual}
\end{figure}

Next, we plot the NMSE for instantaneous CSI estimation as a function of the length of pilot sequence $L$ in Fig. \ref{actual}. As \(L\) increases, it is observed that the NMSEs of both the proposed directional sparsity-aided LS algorithm and the traditional LS algorithm decrease, as expected. However, the proposed directional sparsity-aided LS algorithm achieves significantly higher estimation accuracy compared to the traditional LS algorithm. This is because the proposed LS algorithm exploits the directional sparsity characterized in Stage I (as provided in \eqref{above}), and consequently requires fewer observations than the traditional LS algorithm, which does not exploit this sparsity. The proposed algorithm thus achieves good channel estimation accuracy, even for a small pilot overhead.
Moreover, the NMSE gap between the two considered algorithms increases with $L$.
To achieve the same
estimation accuracy, the proposed  directional sparsity-aided LS algorithm requires 
much lower pilot overhead (i.e., $L$) than the benchmark scheme.
This implies that the effective data rates of users can be significantly increased in Stage III of the proposed protocol.

Next, we conduct simulations to validate the effectiveness of the proposed channel power-based 6DMA position and rotation optimization method. We compare the proposed channel power-based 6DMA PSO algorithm (Algorithm 2) with the following five baseline schemes, where the total number of antennas for each benchmark scheme is set the same as for the proposed scheme. For the proposed and the benchmark schemes, we calculate the sum rate by averaging over 20 channel realizations with the optimized antenna positions and rotations.
\begin{itemize}
    \item Perfect CSI-based continuous 6DMA alternating optimization (AO) \cite{shao20246d}:
	We assume that the instantaneous CSI is perfectly known and AO proposed in \cite{shao20246d} is used to optimize continuous antenna positions and rotations.
		
	\item Perfect CSI-based discrete 6DMA PSO: We assume that the instantaneous CSI is perfectly known and PSO is used to optimize the discrete antenna positions and rotations.
		
	\item Channel power-based discrete 6DMA random-max sampling (RMS): Based on the estimated channel power, we generate  $100$ random samples each consisting of $B$ different  discrete position-rotation pairs and then choose the best sample which yields the highest sum rate.
		
    \item FPA with perfect CSI: We consider a conventional three-sector BS (which can be regarded as a special case  of the 6DMA-BS with $B = 3$ and approximately 
    $\left\lceil \frac{NB}{3} \right\rceil$ antennas on each surface), where each sector antenna surface covers roughly $120^{\circ}$ horizontally. The 3D locations and 3D rotations of all antennas are fixed.
    
    \item Fluid antenna with perfect CSI \cite{new2023fluid}: We consider again a three-sector BS. The rotations of all antennas remain unchanged, while we apply the AO-based algorithm proposed in \cite{shao20246d} to optimize the continuous positions of the antennas within each 2D sector antenna surface.
    
    \item Channel power-based antenna selection: At the BS, we uniformly generate 128 points on a spherical surface with a radius of $\frac{A\sqrt{2}}{2}$, with each point hosting a surface equipped with $N=4$ antennas. Among them, $B=16$ antenna surfaces are selected and activated via the PSO algorithm \cite{con} to maximize the ergodic sum rate, while the remaining antennas are deactivated.
\end{itemize}  
 
\begin{figure}[t!]
	\centering
	\setlength{\abovecaptionskip}{0.cm}
	\includegraphics[width=3.2in]{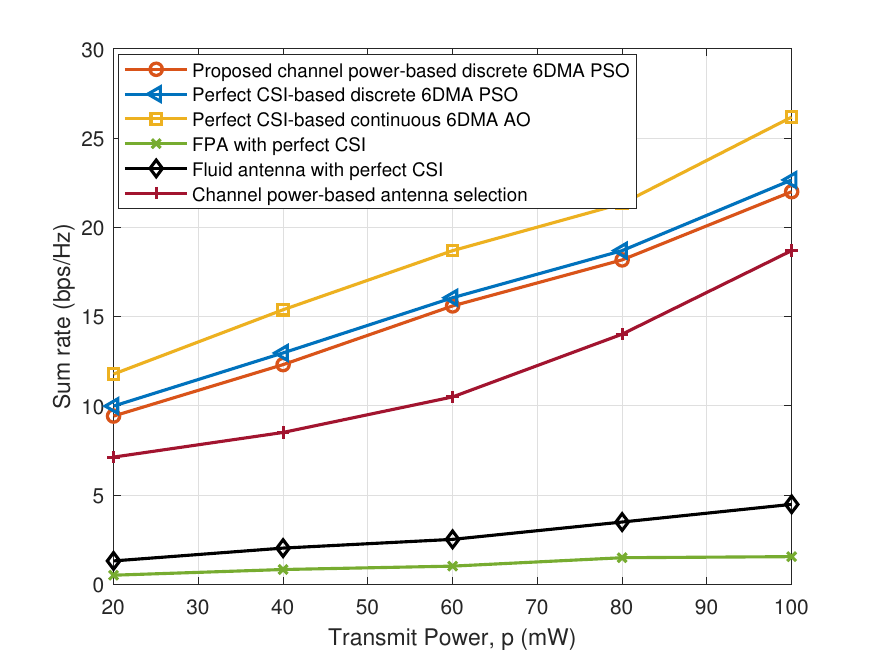}
	\caption{Sum rate versus transmit power for $K=30$ users.}
	\label{pso_power}
\end{figure}

Fig. \ref{pso_power} shows the sum rate achieved by the proposed channel power-based 6DMA PSO algorithm and the baseline schemes versus the user transmit power. It is observed that the sum rate increases with the transmit power. The performance of the proposed algorithm is very close to that of 6DMA PSO with perfect CSI. This is because overcoming deep fading in instantaneous small-scale channels results in minimal gain, and the increase in sum rate is primarily determined by the channel power. This result demonstrates the effectiveness of using average channel power (i.e., statistical CSI instead of instantaneous CSI) for 6DMA position and rotation optimization. 
Furthermore, one can observe that the proposed  algorithm
suffers a performance loss compared to the
AO algorithm with continuous positions and rotations \cite{shao20246d}. This is because the discrete position/rotation adjustments
of the 6DMA surfaces limit their spatial DoFs for adapting to the user channel distribution. 
However, the performance gain realized by continuous adjustments comes at the expense of higher energy consumption and increased control complexity. Therefore, the discrete implementation offers a practical trade-off among performance, cost, and scalability, especially for large-scale 6DMA-BS configurations involving a substantial number of 6DMA surfaces and/or an extensive movement range.
 Furthermore, we see that the proposed channel power-based discrete 6DMA PSO scheme outperforms  antenna selection due to the latter's limited number of possible antenna positions on the spherical surface, which restricts its spatial DoFs for adaptation. Furthermore, antenna selection requires physically
	deploying more candidate antennas, which increases hardware cost and computational complexity compared to 6DMA.

\begin{figure}[t!]
	\centering
	\setlength{\abovecaptionskip}{0.cm}
	\includegraphics[width=3.2in]{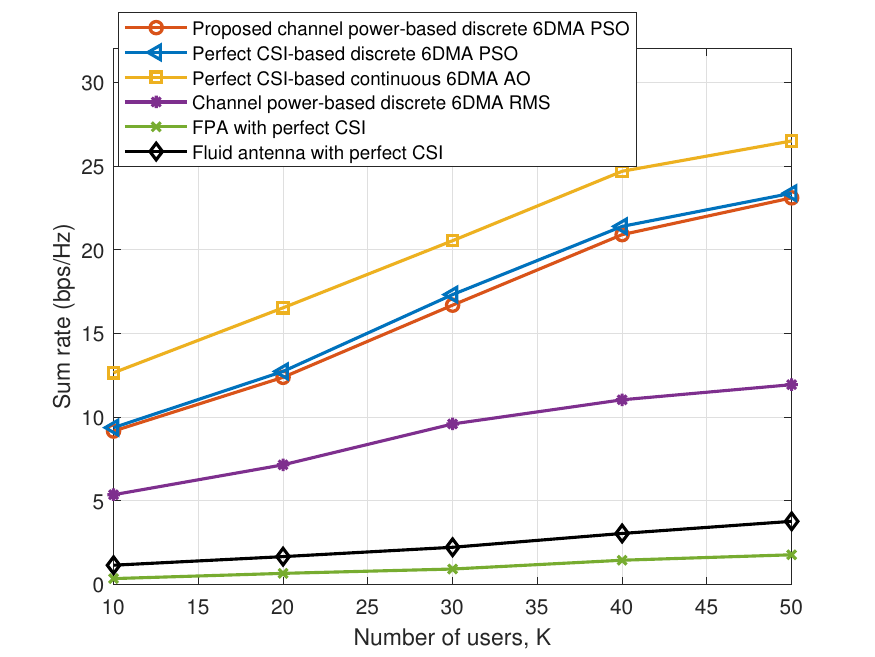}
	\caption{Sum rate versus number of users for transmit power $p=60$ mW.}
	\label{pso_user}
\end{figure}

In Fig. \ref{pso_user}, we compare the sum rate for the considered schemes as a function of the number of users.
First, it is observed that the  proposed channel power-based 6DMA PSO algorithm performs better  than the RMS, defined in our benchmark methods, FPA, and fluid antenna systems. It is also observed that the RMS approach yields the worst 6DMA sum rate performance, which, however, is still significantly better than that of the FPA and fluid antenna systems, even with perfect CSI. This is due to the fact
that the 6DMA scheme has more spatial DoFs and can deploy the antenna resources more efficiently to match the user spatial distribution. On the other hand, the fluid antenna scheme can only adjust the antenna positions within 2D surfaces. Second, it is observed that the performance gaps  become larger as the number of users increases. This suggests that the proposed 6DMA-BS design and algorithms are particularly advantageous for higher network loads and thus more interference-limited scenarios. 

\section{Conclusions}
This paper proposed a distributed 6DMA processing architecture to reduce the centralized  processing complexity by equipping each 6DMA surface with a local LPU. For the new distributed architecture, we presented a practical protocol for operating a BS equipped with 6DMA surfaces. We designed a joint directional sparsity detection and channel power estimation algorithm for statistical CSI acquisition, as well as a directional sparsity-aided instantaneous channel estimation algorithm. Using the estimated  average channel powers, we further designed a statistical channel power-based antenna position and rotation optimization algorithm to maximize the average sum-rate of all users. Extensive simulation results were presented for various practical setups, showing that the proposed channel estimation algorithms achieve higher estimation accuracy than existing schemes while reducing the pilot overhead and also verifying the effectiveness of the proposed practical channel power-based 6DMA position/rotation optimization scheme.

\bibliographystyle{IEEEtran}
\bibliography{fabs}
\end{document}